\documentclass[sigconf]{acmart}

\setcopyright{acmcopyright}
\copyrightyear{2018}
\acmYear{2021}
\acmDOI{10.1145/1122445.1122456}

\acmConference{Conference}{..., 2021}{Virtual}
\acmBooktitle{Conference}
\acmPrice{15.00}
\acmISBN{978-1-4503-XXXX-X/18/06}

\usepackage{lineno,hyperref}
\usepackage{csquotes}

\usepackage{rotating}

\usepackage{multirow}

\usepackage{subcaption}

\usepackage{paralist}

\usepackage{fontawesome}

\usepackage[capitalise,nameinlink]{cleveref}
\crefname{section}{Sect.}{Sect.}
\Crefname{section}{Section}{Sections}
\crefname{figure}{Fig.}{Figs.}
\Crefname{figure}{Figure}{Figures}
\crefname{table}{Tab.}{Tabs.}
\Crefname{table}{Table}{Tables}
\crefname{lstlisting}{List.}{List.}
\Crefname{lstlisting}{Listing}{Listings}

\newcommand{\eg}{e.\,g.,\ }
\newcommand{\ie}{i.\,e.,\ }

\usepackage[inline]{enumitem}

\usepackage{mdframed}
\newmdtheoremenv[innertopmargin=-4pt, innerbottommargin=1pt]{insight}{Insight}

\newcommand{\labeltitle}[1]{\vskip 0.03in \noindent\textbf{#1}} %no full stop

\begin{document}

\title{Demystifying Memory Access Patterns of FPGA-Based Graph Processing Accelerators}

\author{Jonas Dann, Daniel Ritter}
\email{{firstname.lastname}@sap.com}
\affiliation{%
    \institution{SAP SE}
    \city{Walldorf}
    \country{(Germany)}
}

\author{Holger Fröning}
\email{holger.froening@ziti.uni-heidelberg.de}
\affiliation{%
    \institution{Heidelberg University}
    \city{Heidelberg}
    \country{(Germany)}
}

\begin{abstract}
Recent advances in reprogrammable hardware (\eg FPGAs) and memory technology (\eg DDR4, HBM) promise to solve performance problems inherent to graph processing like irregular memory access patterns on traditional hardware (\eg CPU).
While several of these graph accelerators were proposed in recent years, it remains difficult to assess their performance and compare them on common graph workloads and accelerator platforms, due to few open source implementations and excessive implementation effort.

In this work, we build on a simulation environment for graph processing accelerators, to make several existing accelerator approaches comparable.
This allows us to study relevant performance dimensions such as partitioning schemes and memory technology, among others.
The evaluation yields insights into the strengths and weaknesses of current graph processing accelerators along these dimensions, and features a novel in-depth comparison.
\end{abstract}

\maketitle

\vspace{-0.1cm}
\section{Introduction}
\label{sec:intro}
\vspace{-0.1cm}
The irregular memory access and little computational intensity inherent to graph processing cause major performance challenges on traditional hardware (\eg CPUs) due to effects like DRAM row switching and ineffective use of fetched cache lines \cite{journals/corr/abs-1910-09017, journals/corr/abs-2007-07595, LumsdaineGHB07, conf/isca/AhnHYMC15}.
Recent advances in reprogrammable hardware like field programmable gate arrays (FPGAs), and memory technology such as DDR4 and high-bandwidth memory (HBM) promise to accelerate common graph problems (\eg breadth-first search (BFS), PageRank (PR), weakly-connected components (WCC)) \cite{journals/corr/abs-1910-09017, DBLP:journals/corr/abs-2010-13619}.
Especially FPGA-based graph processing accelerators (\eg \cite{conf/IEEEpact/Yao0L0H18, journals/tpds/ZhouKPSW19}) show good results for irregular memory access acceleration due to growing sizes of FPGA on-chip memory and custom memory usage.
However, there are deficiencies in benchmarking of these accelerators due to a multitude of configurations regarding available FPGAs, memory architectures, workloads, input data, and the lack of accepted benchmark standards \cite{journals/corr/abs-2007-07595, DBLP:journals/corr/abs-2010-13619}.
This makes it difficult to assess the implications of different design decisions and optimizations, and --- most importantly --- to compare the proposed accelerators \cite{journals/corr/abs-2007-07595,conf/isca/AhnHYMC15}.

\begin{figure}[bt]
	\centering
	\includegraphics[width=.7\linewidth]{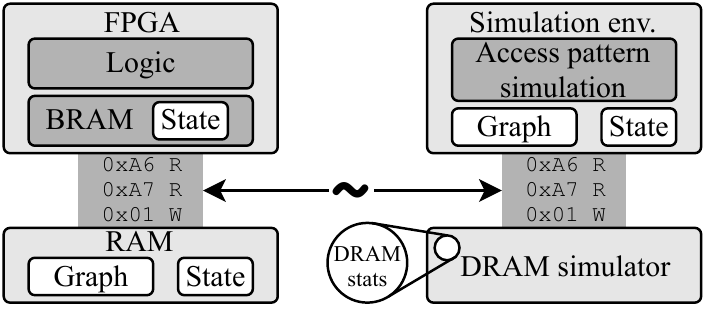}
	\vspace{-0.3cm}
	\caption{DRAM simulator approach (adapted from \cite{DBLP:journals/corr/abs-2010-13619})}
	\label{fig:idea}
\end{figure}

\begin{figure}[bt]
	\centering
	\includegraphics[width=\linewidth]{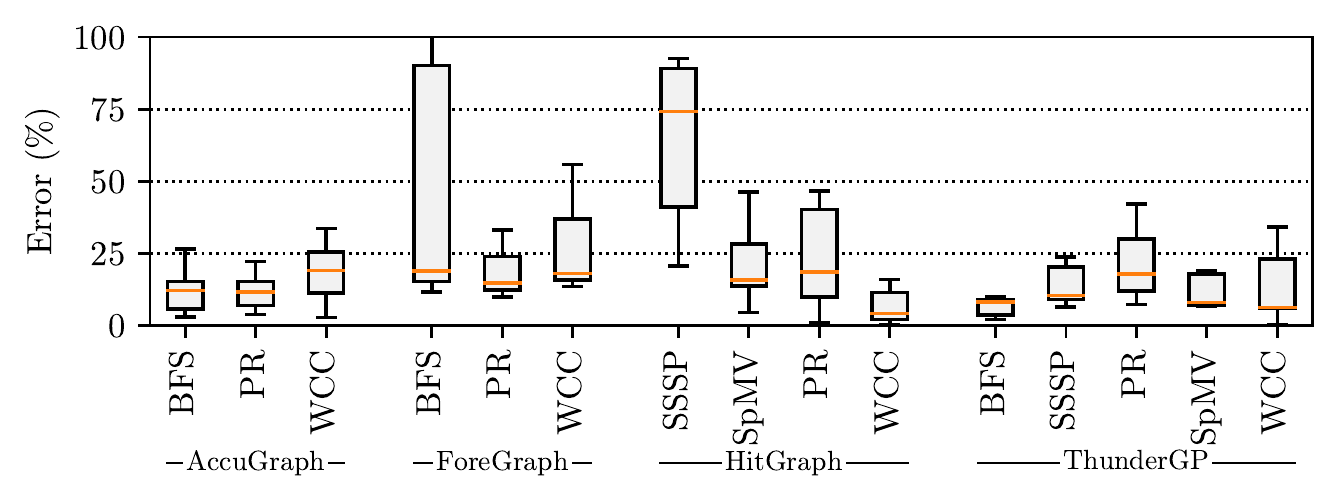}
	\vspace{-0.8cm}
	\caption{Average simulation error by accelerator, workload (from \cite{DBLP:journals/corr/abs-2010-13619} extended by ForeGraph and ThunderGP)\protect\footnotemark}
	\label{fig:reproducibility}
\end{figure}

Recent work \cite{DBLP:journals/corr/abs-2010-13619} introduced a simulator approach (cf. \cref{fig:idea}) that permits to quickly reproduce and compare different graph accelerator solutions in a synthetic, fixed environment. 
Based on the observation that off-chip memory access dominates the overall graph processing accelerator performance, \emph{reproducibility} of results is demonstrated with simulating the approximate request flow based on on-chip (in block RAM (BRAM)) and off-chip state and graph data in the off-chip DRAM.
The results are summarized in \cref{fig:reproducibility}, denoting the simulation error with percentage error $e = \frac{100 \times |s - t|}{t}$, simulation performance $s$ compared to the performance in the respective article $t$ grouped by accelerators (\eg AccuGraph \cite{conf/IEEEpact/Yao0L0H18}, HitGraph \cite{journals/tpds/ZhouKPSW19}) and graph problems.
We get a reasonable mean error of $22.63\%$ with two outliers in BFS on ForeGraph and single-source shortest-paths (SSSP) on HitGraph caused by insufficient specification of root vertices (cf. \cite{DBLP:journals/corr/abs-2010-13619}).

In this work, our objectives are understanding \emph{performance} and \emph{comparability} of graph processing accelerators.
For assessing these objectives, we leverage the simulation approach from \cite{DBLP:journals/corr/abs-2010-13619} to conceptually analyze and compare existing graph processing accelerator approaches along relevant \emph{performance dimensions}: \begin{enumerate*}[label=(\roman*)]
    \item accelerator design decisions,
    \item graph problems,
    \item data set characteristics,
    \item memory technology, and
    \item memory access optimizations.
\end{enumerate*}
Based on criteria like reported performance numbers on commodity hardware and sufficient conceptual details, we choose four state-of-the-art systems found in a recent survey \cite{journals/corr/abs-2007-07595}, namely AccuGraph \cite{conf/IEEEpact/Yao0L0H18}, HitGraph \cite{journals/tpds/ZhouKPSW19}, ForeGraph \cite{conf/fpga/DaiHCXWY17}, and ThunderGP \cite{conf/fpga/ChenTCHWC21} and update \cref{fig:reproducibility} from \cite{DBLP:journals/corr/abs-2010-13619} with simulation error for ForeGraph and ThunderGP.
While AccuGraph and HitGraph are orthogonal approaches representing the currently most relevant paradigms, edge- and vertex-centric graph processing (both with horizontal partitioning) \cite{journals/corr/abs-2007-07595}, ForeGraph is one of the few systems with interval-shard partitioning and compressed edge list, and ThunderGP uses vertical partitioning with a sorted edge list. With that we make the following contributions:
\begin{itemize}
    \itemsep0em
    \item We provide a classification of existing graph processing accelerators and extract their memory access patterns.
    \item We conduct a performance comparison and analysis by comprehensively exploring dimensions (i-iii) and (v).
    \item We analyze graph workloads and accelerator scalability on modern memory like HBM for the first time (dimension (iv)).
\end{itemize}
\footnotetext{Please refer to \cite{DBLP:journals/corr/abs-2010-13619} for details on outliers}

Notably, among other insights, we discover a \emph{trade-off} in the immediate update propagation scheme of AccuGraph and ForeGraph compared to the 2-phase update propagation scheme of HitGraph and ThunderGP.
Additionally, we confirm that modern memory like HBM does not necessarily lead to better performance (cf. \cite{conf/memsys/SchmidtFB16, conf/fccm/WangHZA20}).

\vspace{-0.1cm}
\section{Background}
\label{sec:background}
\vspace{-0.1cm}
In this section, we give a brief background on FPGAs and how they interact with current memory technologies, as well as the simulation environment that this work is based on.

\vspace{-0.1cm}
\subsection{Memory Hierarchy of FPGAs}
\label{sec:ram}
\begin{figure}[bt]
	\centering
	\includegraphics[width=\linewidth]{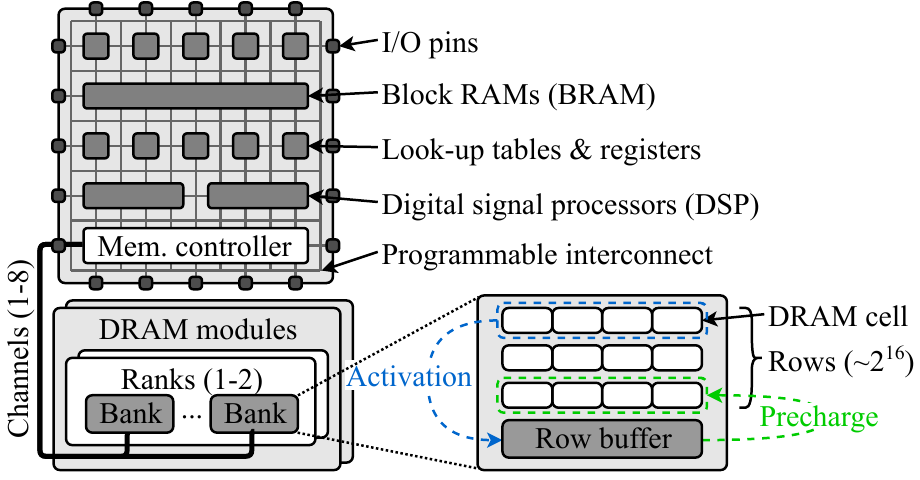}
	\vspace{-0.7cm}
	\caption{FPGA and DRAM (adapted from \cite{journals/cal/KimYM16})}
	\label{fig:ram}
\end{figure}

\vspace{-0.1cm}
As a processor architecture platform, FPGA chips allow for mapping custom digital circuit designs (a set of logic gates and their connections) to a grid of resources (\ie look-up tables, registers) connected with a programmable interconnection network (cf. \cref{fig:ram}).
For frequently used complex functionality like floating point computation FPGAs contain digital signal processors (DSP).
Access to off-chip resources like DRAM and network controllers is possible over I/O pins.
The memory hierarchy of FPGAs is split up into on-chip and off-chip memory.
On-chip, FPGAs implement distributed memory, that is made up of single registers and is mostly used as storage for working values, and block RAM (BRAM) in the form of SRAM memory components for fast storage of data structures.
On modern FPGAs, there is about as much BRAM as there is cache on CPUs (all cache levels combined), but contrary to the fixed cache hierarchies of CPUs, BRAM is finely configurable to the application.

For storage of larger data, DRAM is attached as off-chip memory (\eg DDR3\footnote{JESD79-3 DDR3 SDRAM Standard}, DDR4\footnote{JESD79-4 DDR4 SDRAM Standard}, or HBM\footnote{JESD235D High-Bandwidth Memory (HBM) DRAM Standard}).
Exemplary for DRAM organization, we show the internal organization of DDR3 memory in \cref{fig:ram}.
At the highest level, DRAM is split up into channels each with its own $64$-bit bus to the connected chip.
For each channel, multiple ranks may operate in parallel but on the same bus to increasing memory capacity.
Each rank is then organized as $8$ banks with a row buffer and several thousand rows each.
At the lowest level, data is stored in DRAM cells (with the smallest number of cells addressable at once being called a column).
Read or write requests to data in a bank are served by the row buffer based on three scenarios:
\begin{inparaenum}[\it (1)]
    \item When the addressed row is present in the row buffer, the request is served with low latency (\eg $11$ns).
    \item If the row buffer is empty, activating the addressed row adds additional latency (\eg $11$ns).
    \item However, if the row buffer currently contains a different row, the present row has to be pre-charged (\eg $11$ns), before the addressed row can be activated and the request served.
\end{inparaenum}
As a reference point, a last level cache miss on current Skylake Intel CPUs takes $17$ns at $2.6$GHz clock.
Additionally, there is a minimum latency between switching rows (\eg $28$ns).
Thus, for high performance, row switching should be avoided as much as possible.
To achieve good performance, current DRAM additionally employ $8$n prefetching (meaning $8$ bursts).
Thus, 64 bytes are returned for each request which we call a cache line in the following.
DDR4 doubles total number of banks at the cost of added latency due to another hierarchy level called bank groups, which group two to four banks.
HBM as a new stacked memory technology introduces very high bandwidth in a small package.
The single channels have double as many banks ($16$) as DDR3 with half the prefetch ($4$n) which however transport double the data per cycle ($128$-bit).
Additionally, HBM has smaller row buffers and can have many more channels in a confined space.

\vspace{-0.1cm}
\subsection{Memory Access Simulation Environment}
\label{sec:environment}
\vspace{-0.1cm}
Since memory access is the dominating factor in graph processing, the necessity of cycle-accurate simulation of on-chip data flow is relaxed and only the off-chip memory access pattern is modeled in the simulation environment \cite{DBLP:journals/corr/abs-2010-13619}.
This means modelling request types, addressing, volume, and ordering.
Request type modelling is trivial since it is clear when requests read and write data.
For request addressing, we assume that the different data structures lie adjacent in memory as plain arrays.
We generate memory addresses according to this memory layout and the width of the array types in bytes.
Request volume modelling is mostly based on the size $n$ of the vertex set, the size $m$ of the edge set, average degree $deg$, and partition number $k$ of a given graph.
Lastly, we only simulate request ordering through mandatory control flow caused by data dependencies of requests.
We assume that computations and on-chip memory accesses are instantaneous by default.
The simulation implementations are further based on memory access abstractions that we introduce in \cref{sec:selected} as needed.

A DRAM simulator is an integral part of the simulation environment (cf. \cref{fig:idea}).
For our purposes we need a DRAM simulator that supports at least DDR3 (for HitGraph \cite{journals/tpds/ZhouKPSW19}) and DDR4 (for AccuGraph \cite{conf/IEEEpact/Yao0L0H18}, ForeGraph \cite{conf/fpga/DaiHCXWY17}, and ThunderGP \cite{conf/fpga/ChenTCHWC21}).
We chose Ramulator \cite{journals/cal/KimYM16} for this work over other alternatives like DRAMSim2 \cite{journals/cal/RosenfeldCJ11} or USIMM \cite{ChatterjeeBSPUSSAC12}.
To the best of our knowledge, it is the only DRAM simulator which supports both DDR3 and DDR4 (among many others like LPDDR3/4 and HBM).
The relevant Ramulator configuration parameters for this work are:
\begin{inparaenum}[\it (a)] 
    \item DRAM standard,
    \item Number of channels,
    \item Number of ranks,
    \item DRAM speed specification, and
    \item DRAM organization.
\end{inparaenum}

\vspace{-0.1cm}
\section{Graph Processing Accelerators}
\label{sec:accelerators}
\vspace{-0.1cm}
In this section, we introduce basics of graph processing accelerators and thereafter motivate the selection of accelerators for this work and show their implementation in the simulation environment.

\vspace{-0.1cm}
\subsection{Partition, Iteration, \& Update Schemes}
\label{sec:graph_basics}
\vspace{-0.1cm}
Graph processing accelerators utilize two dimensions of graph partitioning: horizontal and vertical.
Horizontal partitioning means dividing up the vertex set into equal intervals and letting each partition contain the outgoing edges of one interval.
Examples are AccuGraph which uses a horizontally partitioned compressed sparse row (CSR) format of the inverted edges and HitGraph which uses a horizontally partitioned edge list.
Vertical partitioning divides the vertex set into intervals like horizontal partitioning but each partition contains the incoming edges of its interval.
ThunderGP uses a vertically partitioned edge list.
As a third partitioning approach, interval-shard partitioning \cite{conf/usenix/ZhuHC15} employs both vertical and horizontal partitioning at once and is used by ForeGraph.

Depending on the underlying graph data structure, graphs are processed based on two fundamentally different approaches: edge- and vertex-centric graph processing.
Edge-centric systems (\eg ForeGraph, HitGraph, and ThunderGP) iterate over the edges as primitives of the graph on an underlying edge list.
Vertex-centric systems iterate over the vertices and their neighbors as primitives of the graph on an underlying adjacency list (\eg CSR).
For the vertex-centric approach there further is a distinction into push- and pull-based data flow.
A push-based data flow denotes that values are pushed along the forward direction of edges to update neighboring vertices.
A pull-based data flow (\eg AccuGraph) denotes that values are pulled along the inverse direction of edges from neighboring vertices to update the current vertex.

Lastly, there are currently three update propagation schemes.
Immediate update propagation directly applies updates to the working value set when they are produced, 2-phase update propagation collects all updates in memory and applies them in a separate phase of each iteration, and level-synchronous update propagation, specifically for BFS, maintains a frontier of vertices with updates in each iteration only on their corresponding values.

\vspace{-0.1cm}
\subsection{Selected Accelerators}
\label{sec:selected}
\vspace{-0.1cm}
\begin{table*}[t]
\footnotesize
\centering
\begin{tabular}{l|l l|l l l l l}
    Identifier & Supported graph problems & System & Iter. scheme & Flow & Partitioning & Binary rep. & Update prop. \\
    \hline
    \hline
    \textbf{ForeGraph} \cite{conf/fpga/DaiHCXWY17} & BFS, PR, WCC & Simulation & Edge-centric & n/a & Interval-shard & Compr. edge list & Immediate \\
    \textbf{HitGraph} \cite{journals/tpds/ZhouKPSW19} & BFS, PR, WCC, SSSP, SpMV & FPGA & Edge-centric & n/a & Horizontal & Sorted edge list & 2-phase \\
    \textbf{ThunderGP} \cite{conf/fpga/ChenTCHWC21} & BFS, PR, WCC, SSSP, SpMV, ... & FPGA & Edge-centric & n/a & Vertical & Sorted edge list & 2-phase \\
    \hline
    \textbf{AccuGraph} \cite{conf/IEEEpact/Yao0L0H18} & BFS, PR, WCC & FPGA & Vertex-centric & Pull & Horizontal & in-CSR & Immediate \\
    Betkaoui et al. \cite{conf/asap/BetkaouiWTL12} & BFS & Convey HC-2 & Vertex-centric & Push & Horizontal & CSR & Level-synch. \\
    CyGraph \cite{conf/ipps/AttiaJTJZ14} & BFS & Convey HC-2 & Vertex-centric & Push  & None & CSR (in-place) & Level-synch. \\
    Ayupov et al. \cite{journals/tcad/AyupovYOKBO18} & BFS, PR, WCC & Simulation & Vertex-centric & Pull & None & in-CSR & Immediate \\
    TorusBFS \cite{journals/iracst/LeiRG15} & BFS & Convey HC-2 & Vertex-centric & Push & None & CSR & Level-synch. \\
    \hline
    Yang et al. \cite{journals/fcsc/YangZGJ20} & BFS, PR, WCC & FPGA & Hybrid & n/a & Vertical & EBs \& CSR & 2-phase \\
    Zhang et al. \cite{conf/fpga/ZhangL18} & BFS & HMC FPGA & Vertex-centric & Hybrid & None & CSR & Level-synch. \\
\end{tabular}

\medskip
Iter. scheme: Iteration scheme; Binary rep.: Binary representation; Update prop.: Update propagation; n/a: Not applicable; EBs: Edge blocks

\caption{FPGA-accelerated graph systems in context of most relevant design and implementation decisions}
\label{tab:decision}
\end{table*}
For this work, we selected four graph processing accelerators from the list of accelerators in \cite{journals/corr/abs-2007-07595}, based on if
\begin{enumerate*}[label=(\roman*)]
    \item they run on commodity FPGAs,
    \item the vertex set is not required to fit into on-chip memory, and
    \item the respective article provides enough detail to model the memory access pattern.
\end{enumerate*}
The accelerators fitting all but one of these criteria can be found in \cref{tab:decision} and the ones we chose to include in this work are highlighted in bold.
We excluded the accelerators based on the Convey HC-2 and HMC systems because they are not implementable with commercially available FPGAs anymore and not reproducible.
Additionally, only supporting BFS restricts their usefulness.
Furthermore, we excluded the systems by Ayupov et al. \cite{journals/tcad/AyupovYOKBO18} and Yang et al. \cite{journals/fcsc/YangZGJ20} because they did not provide sufficient detail to reproduce the results with the simulation environment.
The set of accelerators we choose for this work represents the currently highest performing graph processing accelerators and all currently applied partitioning and iteration schemes.
In the following, we introduce how they can be implemented in the simulation environment in alphabetical order.
For more detail than we provide here on the notation and implementation aspects, we refer to \cite{DBLP:journals/corr/abs-2010-13619}.

\vspace{-0.1cm}
\subsubsection{AccuGraph}
\label{sec:accugraph}
\begin{figure}[bt]
	\centering
	\includegraphics[width=\linewidth]{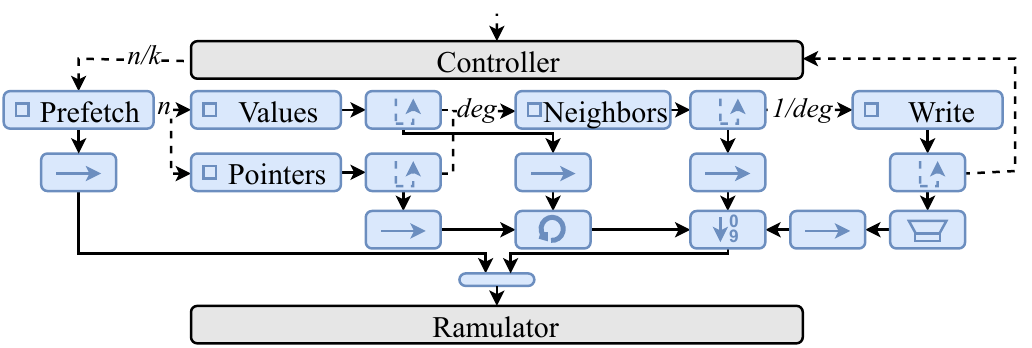}
	\vspace{-0.7cm}
	\caption{AccuGraph (adapted from \cite{DBLP:journals/corr/abs-2010-13619})}
	\label{fig:accugraph}
\end{figure}
AccuGraph \cite{conf/IEEEpact/Yao0L0H18} proposes a flexible accumulator based on a modified prefix-adder able to produce and merge updates to multiple vertices per cycle.
\cref{fig:accugraph} shows the request and control flow modelling of AccuGraph.
The controller is triggered and iterates over the graph until there are no more changes in the previous iteration.
Each iteration triggers processing of all partitions.
Processing of each partition starts with triggering the prefetch request \emph{producer} which prefetches the partitions $\frac{n}{k}$ vertex values (with $n = |V|$ and the partition count $k$) sequentially which is passed through a \emph{cache line} memory access abstraction merging adjacent requests to the same cache line into one.
Thereafter, values and pointers of all destination vertices are fetched sequentially.
Both of those request streams are annotated with \emph{callbacks} which return control flow for each served request.
Those two request streams are merged \emph{round-robin}, because a value is only useful with the associated pointers.
In parallel, neighbors are read from memory sequentially, annotated with their value from the prefetched vertex values, and one edge is materialized for each neighbor with its corresponding source vertex.
The aforementioned accumulator thereafter produces updates for each edge.
If the source vertex value changes, the result is written back to off-chip memory (unchanged values are filtered by the \emph{filter} memory access abstraction).
All of these request streams are merged by \emph{priority} with write request taking the highest priority and neighbors the second highest.

\vspace{-0.1cm}
\subsubsection{ForeGraph}
\label{sec:foregraph}
\begin{figure}[bt]
	\centering
	\includegraphics[width=.9\linewidth]{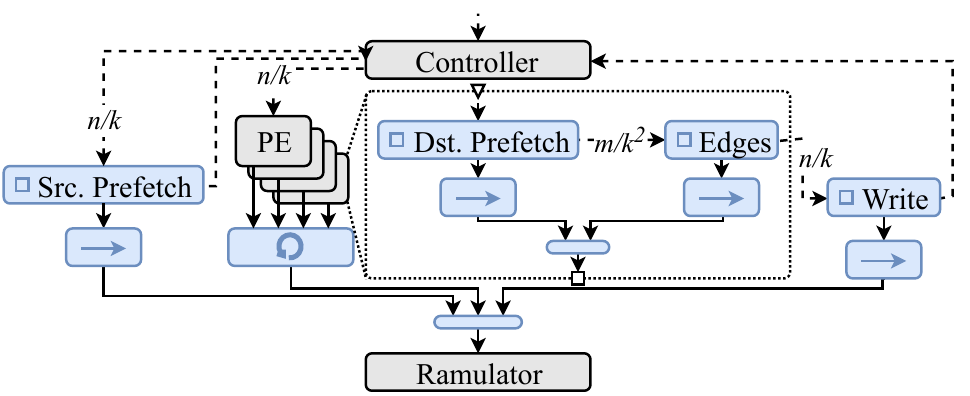}
	\vspace{-0.3cm}
	\caption{ForeGraph}
	\label{fig:foregraph}
\end{figure}

ForeGraph \cite{conf/fpga/DaiHCXWY17} stores the edges of the shards as compressed 32-bit edges with two 16-bit vertex identifiers each.
This is possible due to the interval size being limited to $65,536$ vertices.
In each iteration, ForeGraph prefetches the source intervals one after another and for each source interval processes its corresponding shards by additionally prefetching the destination interval and sequentially reading and processing the edges (\cref{fig:foregraph}).
This results in purely sequential off-chip requests with all random vertex value requests during edge processing being served by caches on-chip. 
After a shard has been processed, the destination interval is sequentially written back to off-chip memory.
To achieve competitive performance, ForeGraph instantiates $p$ processing elements (PE).
These each work on their own set of source intervals and share the memory access in round-robin fashion.

\vspace{-0.1cm}
\subsubsection{HitGraph}
\label{sec:Hitgraph}
\begin{figure}[bt]
	\centering
	\includegraphics[width=\linewidth]{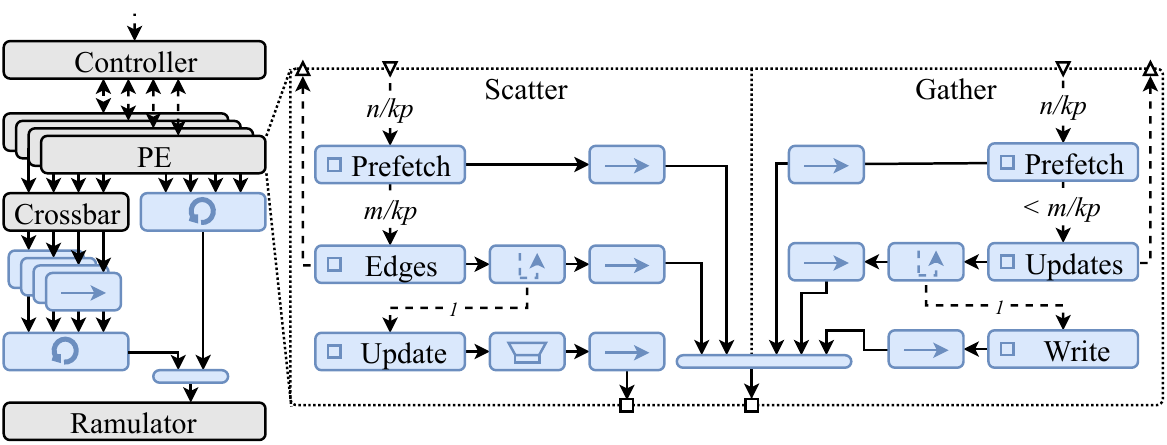}
	\vspace{-0.7cm}
	\caption{HitGraph (adapted from \cite{DBLP:journals/corr/abs-2010-13619})}
	\label{fig:hitgraph}
\end{figure}
HitGraph \cite{journals/tpds/ZhouKPSW19} execution starts with triggering a controller that itself triggers iterations of edge-centric processing until there were no value changes in the previous iteration (\cref{fig:hitgraph}).
In each iteration, the controller first schedules all $k$ partitions for the scatter phase (producing updates), before scheduling all partitions to the gather phase (applying updates).
Partitions are assigned beforehand to channels of the memory (four channels in \cite{journals/tpds/ZhouKPSW19}) and there are $p$ processing elements (PE), one for each channel.

The scatter phase starts by prefetching the $\frac{n}{kp}$ values of the current partition.
After all requests are produced, the prefetch step triggers the edge reading step that reads all ${\sim}\frac{m}{kp}$ edges of the partition.
For each edge request, we attach a callback that triggers producing an update request.
The target address depends on its destination vertex that can be part of any of the $k$ partitions.
Thus, there is a crossbar (unique to this accelerator) that routes each update request to a cache line access abstraction for each partition to sequentially write into a partition-specific update queue.

Similar to scatter, the gather phase starts with prefetching the $\frac{n}{kp}$ vertex values of the current partition sequentially.
After value requests have been produced, the prefetch producer triggers the update producer, which sequentially reads the update queue written by the scatter phase before.
For each update we register a callback that triggers the value write.
The value writes are not necessarily sequential, but especially for iterations where a lot of values are written, there is a lot of locality.

All request streams in each PE are merged directly into one stream without any specific merging logic, since only one producer is producing requests at a time.
Since all PEs are working on independent channels and Ramulator only offers one endpoint for all channels combined, we employ a round-robin merge of the PE requests in order not to starve any channel.

\vspace{-0.1cm}
\subsubsection{ThunderGP}
\label{sec:thundergp}
\begin{figure}[bt]
	\centering
	\includegraphics[width=\linewidth]{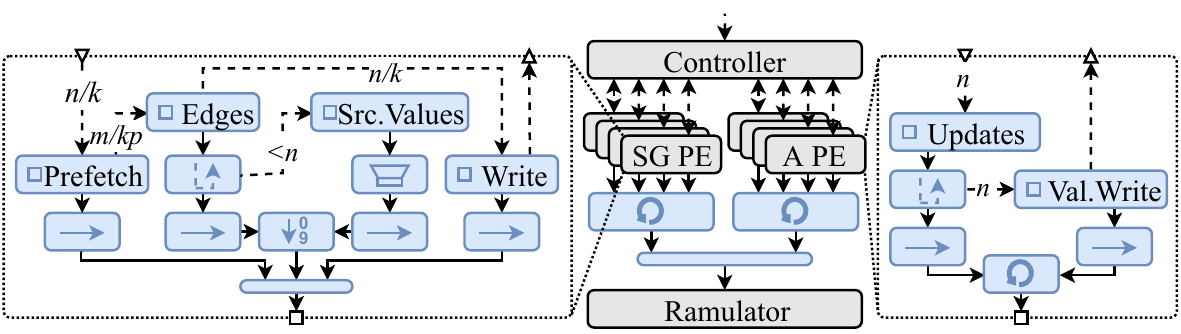}
	\vspace{-0.7cm}
	\caption{ThunderGP}
	\label{fig:thundergp}
\end{figure}
Like HitGraph, ThunderGP \cite{conf/fpl/ChenBCHHWC19} is based on a 2-phase update propagation scheme based on edge-centric iteration over the input graph.
The graph is vertically partitioned into $k$ partitions and each partition is split up into $p$ chunks where $p$ is equal to the number of memory channels.
Each memory channel contains the whole vertex value set of the graph, its corresponding chunk of each partition and an update set.
For each iteration, a scatter-gather phase (SG PE) is applied for each partition before an apply phase (A PE) is executed for each partition (\cref{fig:thundergp}).
The scatter-gather phase starts by prefetching the partitions destination vertex value set into BRAM sequentially.
Upon finalization, this triggers sequential edge reading.
Each edge callback triggers loading its source vertex value. 
Since edge lists are sorted by source vertex, this is semi-sequential and a vertex value buffer filters duplicate source vertex value requests.
When edge reading is finished, the updated values are written back to memory.
The apply stage reads all updates sequentially and combines the updates produced by all channels in the previous phase into one value per vertex which is written back to all channels.
This step may cause many duplicate value reads and writes.

\vspace{-0.1cm}
\section{Accelerator comparison}
\label{sec:comparison}
\vspace{-0.1cm}
So far, graph accelerators were compared by looking at their absolute reported performance numbers, and thus making comparisons based on different FPGAs, memory architectures, number of memory channels, and consequently off-chip memory bandwidth available (\eg \cite{journals/tpds/ZhouKPSW19,conf/IEEEpact/Yao0L0H18}).
In this section, we collect graph data sets used by the four accelerators from \cref{sec:accelerators} and their simulation environment configuration parameters to compare them along performance dimensions relevant for graph processing:
\begin{enumerate*}[label=(\roman*)]
    \item accelerator design decisions and
    \item graph problems,
    \item data set characteristics,
    \item memory technology, and
    \item memory access optimizations.
\end{enumerate*}

\vspace{-0.1cm}
\subsection{Setup: Data Sets \& Simulation Environment}
\label{sec:configuration}
\vspace{-0.1cm}
\begin{table}[bt]
\footnotesize
\centering
\begin{tabular}{l r r c c r r r}
 Name & $|V|$ & $|E|$ & Dir. & Degs. & $D_{avg}$ & \o & SCC \\
 \hline
 \hline
 twitter (tw) & $41.7$M & $1,468.4$M & \faThumbsOUp & \includegraphics[height=1em]{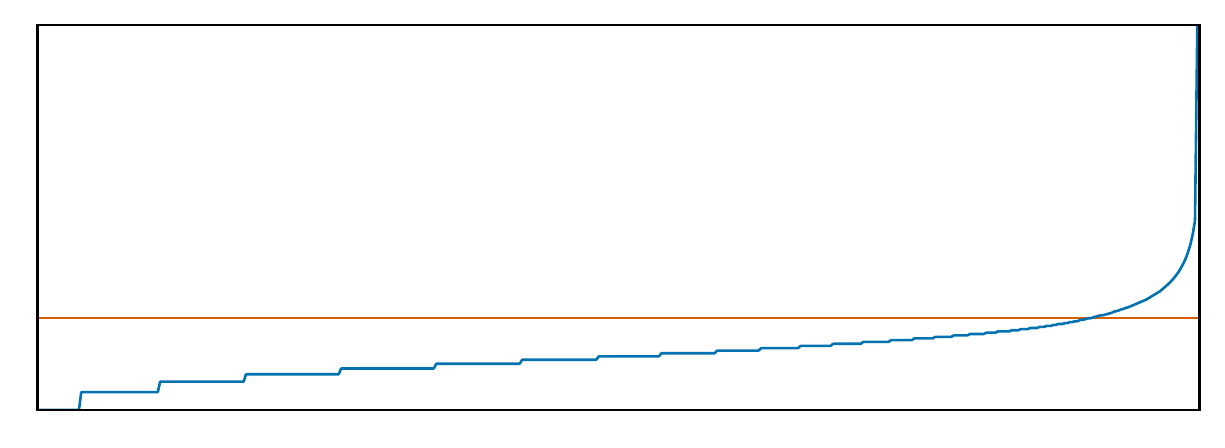} & $35.25$ & $75$ & $0.80$ \\
 live-journal (lj) & $4.8$M & $69.0$M & \faThumbsOUp & \includegraphics[height=1em]{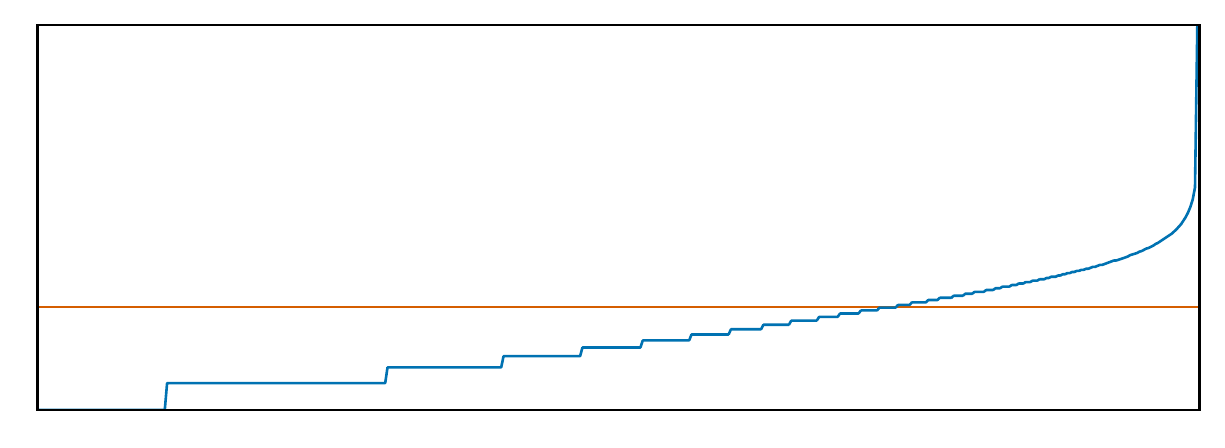} & $14.23$ & $20$ & $0.79$ \\
 orkut (or) & $3.1$M & $117.2$M & \faThumbsDown & \includegraphics[height=1em]{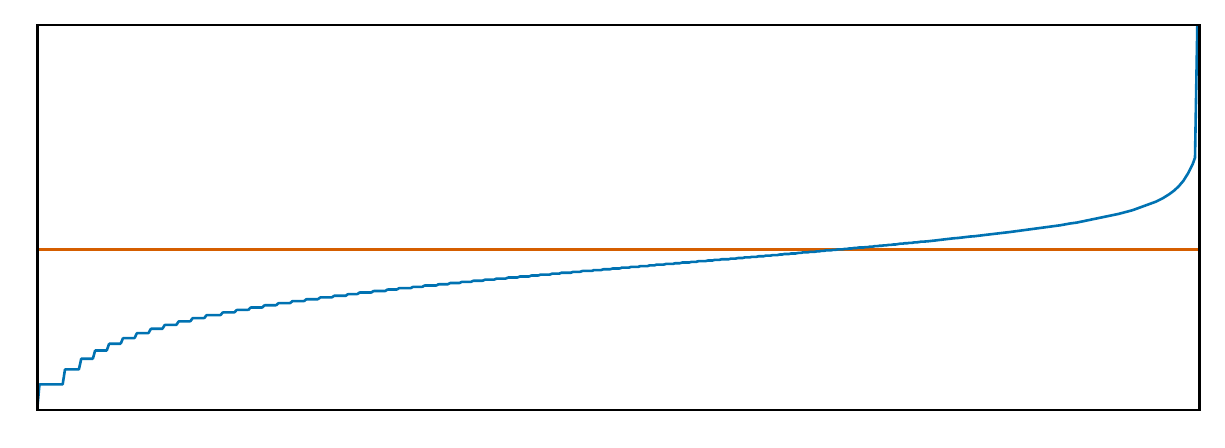} & $76.28$ & $9$ & $1.00$ \\
 wiki-talk (wt) & $2.4$M & $5.0$M & \faThumbsOUp & \includegraphics[height=1em]{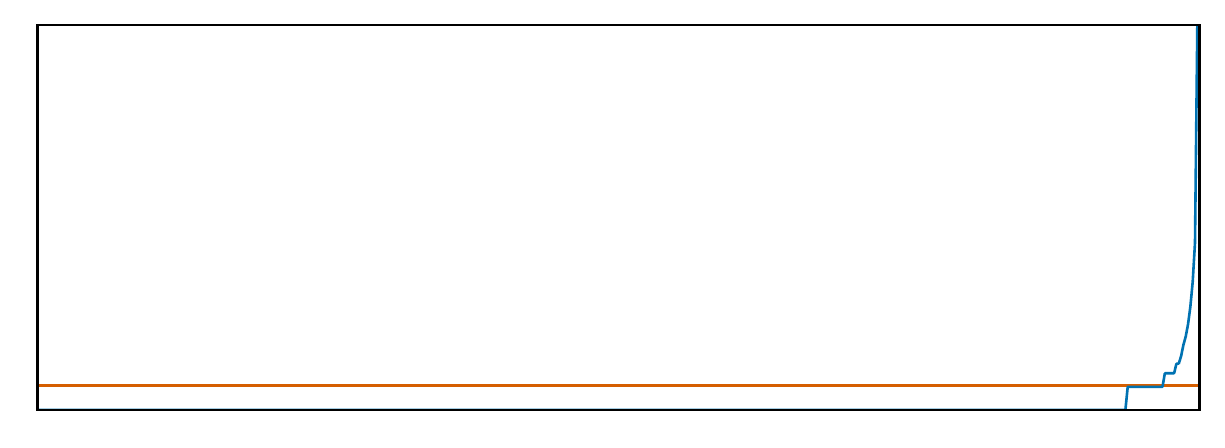} & $2.10$ & $11$ & $0.05$ \\
 pokec (pk) & $1.6$M & $30.6$M & \faThumbsDown & \includegraphics[height=1em]{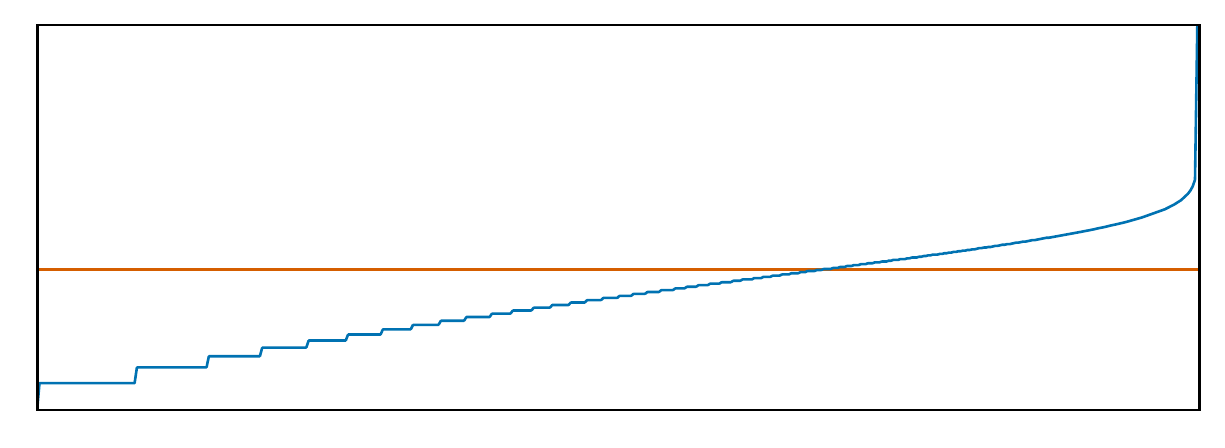} & $37.51$ & $14$ & $1.00$ \\
 youtube (yt) & $1.2$M & $3.0$M & \faThumbsDown & \includegraphics[height=1em]{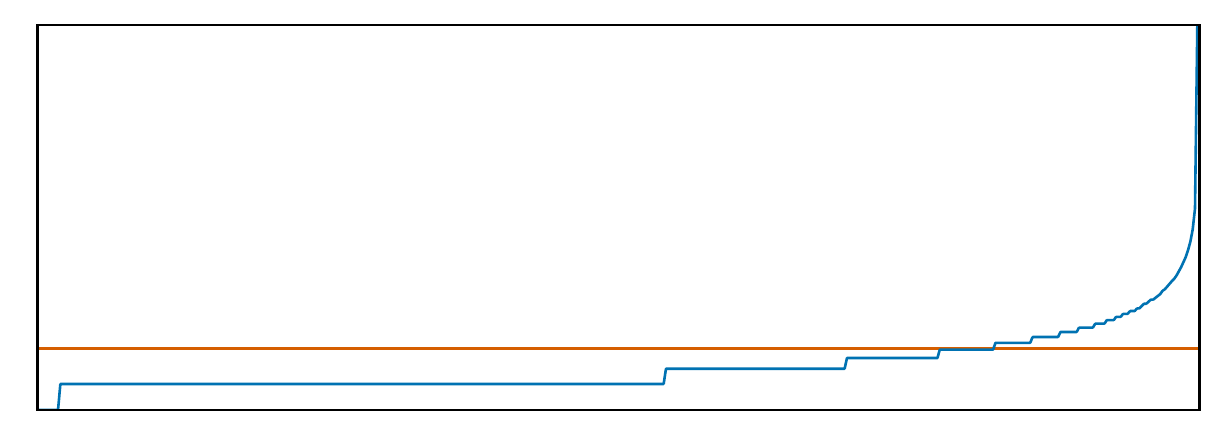} & $5.16$ & $20$ & $0.98$ \\
 dblp (db) & $426.0$K & $1.0$M & \faThumbsDown & \includegraphics[height=1em]{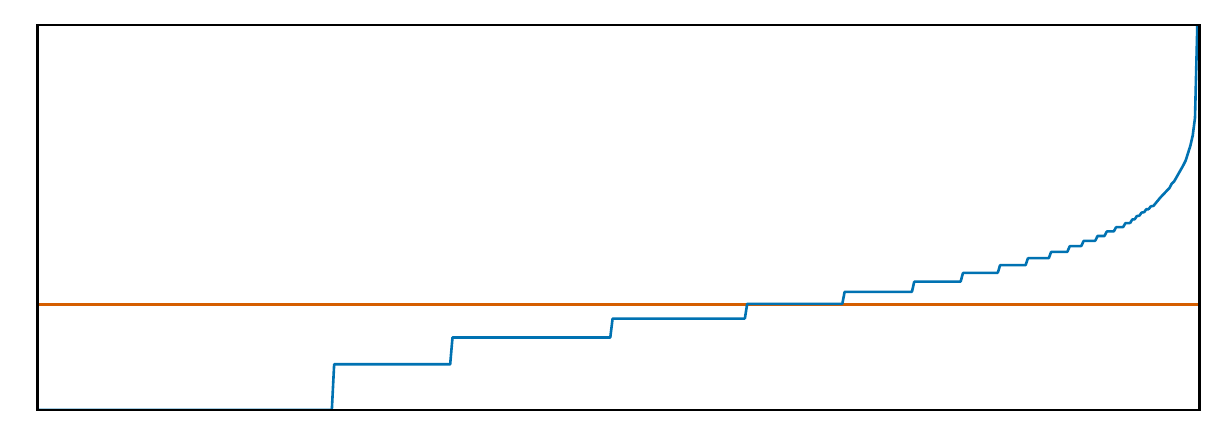} & $4.93$ & $21$ & $0.74$ \\
 slashdot (sd) & $82.2$K & $948.4$K & \faThumbsOUp & \includegraphics[height=1em]{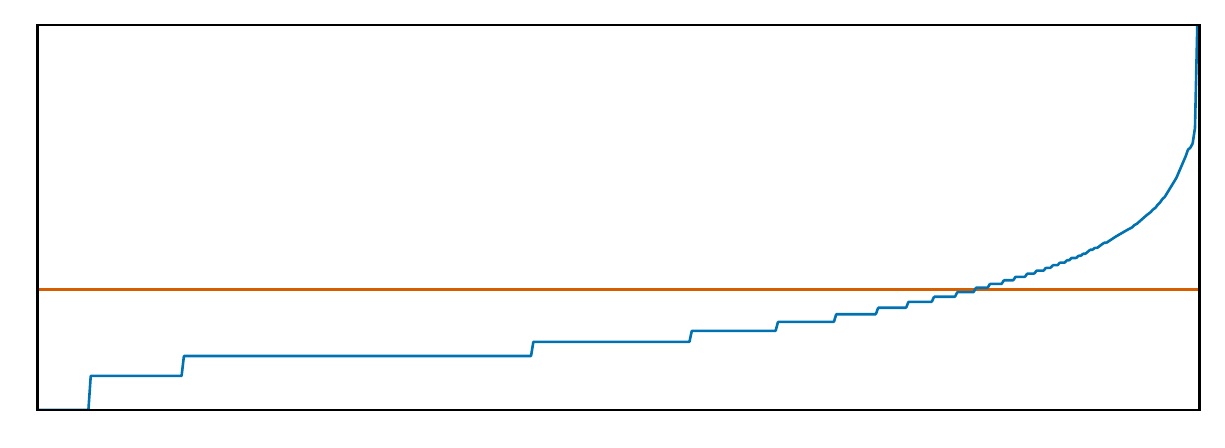} & $11.54$ & $13$ & $0.87$ \\
 \hline
 roadnet-ca (rd) & $2.0$M & $2.8$M & \faThumbsDown & \includegraphics[height=1em]{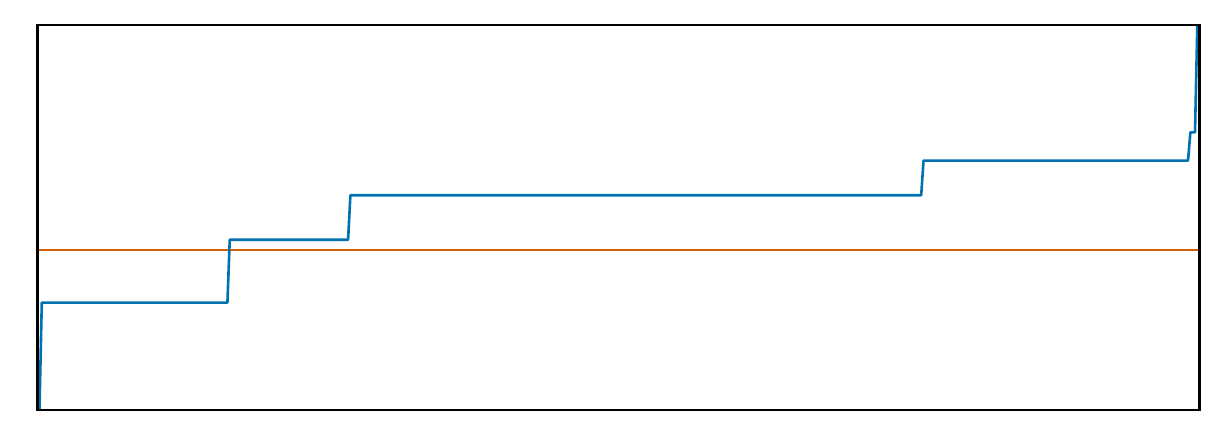} & $2.81$ & $849$ & $0.99$ \\
 berk-stan (bk) & $685.2$K & $7.6$M & \faThumbsOUp & \includegraphics[height=1em]{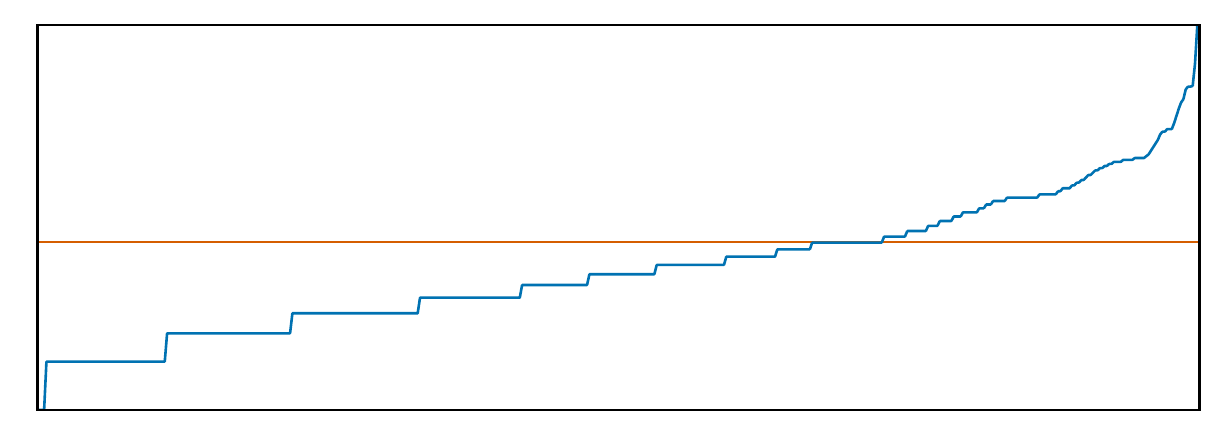} & $11.09$ & $714$ & $0.49$ \\
 \hline
 rmat-24-16 (r24) & $16.8$M & $268.4$M & \faThumbsOUp & \includegraphics[height=1em]{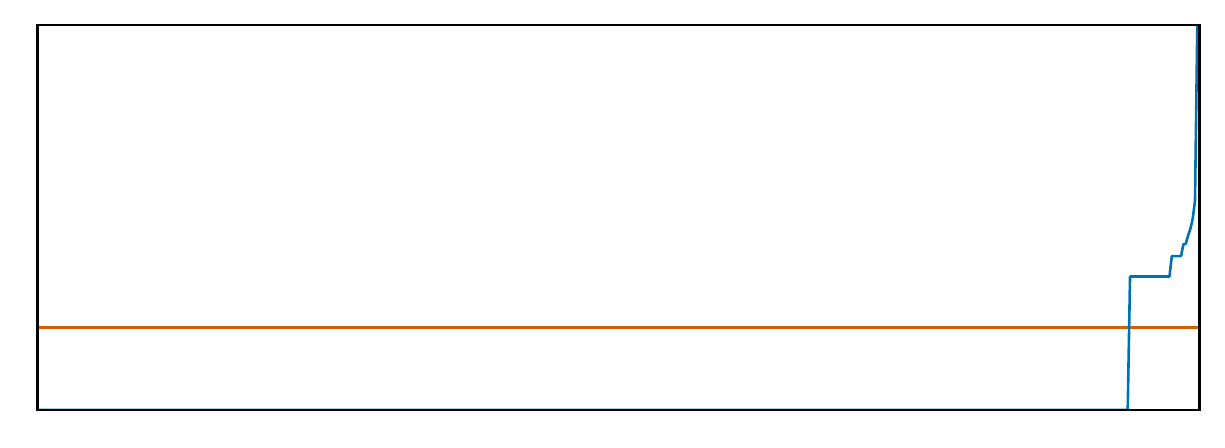} & $16.00$ & $19$ & $0.02$ \\
 rmat-21-86 (r21) & $2.1$M & $180.4$M & \faThumbsOUp & \includegraphics[height=1em]{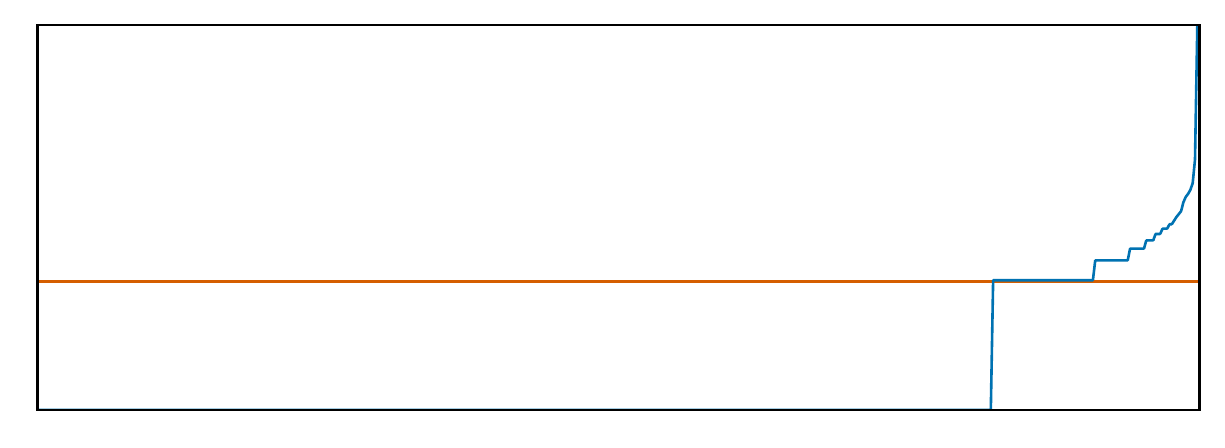} & $86.00$ & $14$ & $0.10$ \\
\end{tabular}

\medskip
Dir.: Directed; Degs.: Degree distribution on log. scale; SCC: Ratio of vertices in the largest strongly-connected component to $n$; \faThumbsOUp: yes, \faThumbsDown: no

\caption{Graphs used often by systems in \cref{tab:decision} (real-world graphs from SNAP \cite{LeskovecK14}; Graph500 generator for R-MAT)}
\label{tab:graphs}
\end{table}
Graph data sets that are often used to benchmark the systems in \cref{tab:decision} are listed in \cref{tab:graphs}.
This selection represents the most important graphs, currently considered, found by a recent survey \cite{journals/corr/abs-2007-07595}.
Two important aspects when working with these graphs are their directedness and the choice of root vertices\footnote{Root vertices: tw - $2748769$; lj - $772860$; or - $1386825$; wt - $17540$; pk - $315318$; yt - $140289$; db - $9799$; sd - $30279$; rd - $1166467$; bk - $546279$; r24 - $535262$; r21 - $74764$} (\eg for BFS or SSSP), because they can have a significant impact on performance.

Regarding data types, we use 32-bit data types for all vertex identifiers, CSR pointers, and values (integer or float).
The only exception is ForeGraph which can use 16-bit vertex identifiers due to its interval-shard partitioning.
An unweighted edge is always two vertex identifiers wide and a weighted edge is an additional 32 bits wider due to the attached edge weight. 
This is sensible for all accelerators we encountered, since there are no excessively large benchmark graphs or requirements on more precision.

\begin{table}[bt]
\footnotesize
\centering
\begin{tabular}{l|l r r r r r r}
    Identifier & Type & Chan. & Ranks & Data rate & BW & Size & RBS \\
    \hline
    \hline
    AccuGraph & DDR4 & $1$ & $1$ & $2400$MT/s & $19.2$GB/s & $2$Gb & $8$KB \\
    ForeGraph & DDR4 & $1$ & $1$ & $2400$MT/s & $19.2$GB/s & $4$Gb & $8$KB \\
    HitGraph  & DDR3 & $4$ & $2$ & $1600$MT/s & $12.8$GB/s & $8$Gb & $8$KB \\
    ThunderGP & DDR4 & $4$ & $1$ & $2400$MT/s & $19.2$GB/s & $16$Gb & $8$KB \\
    \hline
    Default & DDR4 & $1-4$ & $1$ & $2400$MT/s & $19.2$GB/s & $16$Gb & $8$KB \\
    DDR3    & DDR3 & $1-4$ & $1$ & $2133$MT/s & $17.1$GB/s & $8$Gb & $8$KB \\
    HBM     & HBM  & $1-8$ & n/a & $1000$MT/s & $16.0$GB/s & $4$Gb & $2$KB \\
\end{tabular}

\medskip
Chan.: Channels; BW: Bandwidth / Chan.; RBS: Row buffer size; n/a: not applicable

\caption{DRAM configurations}
\label{tab:dram}
\end{table}

\cref{tab:dram} shows the DRAM configurations used in the respective paper of the selected accelerators as well as the DRAM configurations we use in this work.
The default is the DDR4 configuration, since this is the most common in the systems the selected accelerators run on.
The DDR3 and HBM configurations are used to compare performance on different memory technologies in \cref{sec:dram}.
By default, we use one channel, but also do a scale test for HitGraph and ThunderGP due to their support for multiple channels.

We consider the five graph problems breadth first search (BFS), PageRank (PR), weakly connected components (WCC), sparse matrix-vector multiplication (SpMV), and single-source shortest-paths (SSSP).
However, SpMV and SSSP require edge weights which only HitGraph and ThunderGP support.

We use Ramulator\footnote{Ramulator, visited 3/21: \url{https://github.com/CMU-SAFARI/ramulator}} commit \texttt{dd326} and add a function to flush the stats for multiple consecutive runs.
To compile Ramulator and the simulation environment we use \texttt{clang++} 5.0.1.
For Ramulator with C++11 and for the simulation environment with C++17 and \texttt{-D\_FILE\_OFFSET\_BITS=64} to be able to read files larger than 2GB.

In subsequent performance measurements, we use the Graph500 benchmark's MTEPS definition, which specifies MTEPS as $|E| / t_{exec}$, where $ t_{exec}$ denotes the execution time.
Notably, this is different to the measure that most graph processing accelerator articles report (\ie total number of edges read during execution divided by execution time), which we call MREPS.
MREPS do not normalize the runtime to graph size but rather denote raw edge processing performance.
For both MTEPS and MREPS higher is better.

\vspace{-0.1cm}
\subsection{Design Decisions \& Graph Problems}
\label{sec:observation}
\vspace{-0.1cm}
\begin{figure*}[bt]
	\centering
	\includegraphics[width=\linewidth]{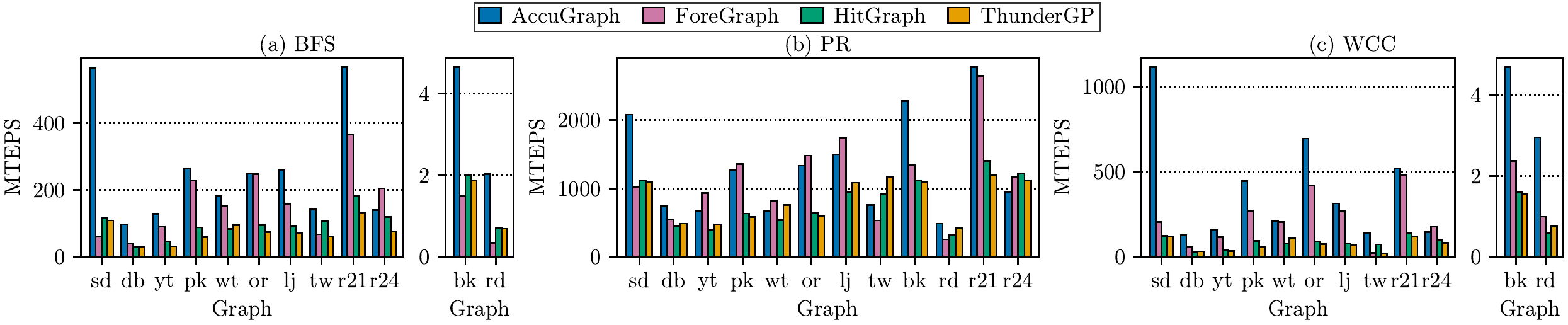}
	\vspace{-0.7cm}
	\caption{Comparison of graph processing accelerators grouped by graphs for BFS, PR, and WCC (DDR4, single-channel)\protect\footnotemark}
	\label{fig:comparison}
\end{figure*}
\begin{figure}[bt]
	\centering
	\includegraphics[width=\linewidth]{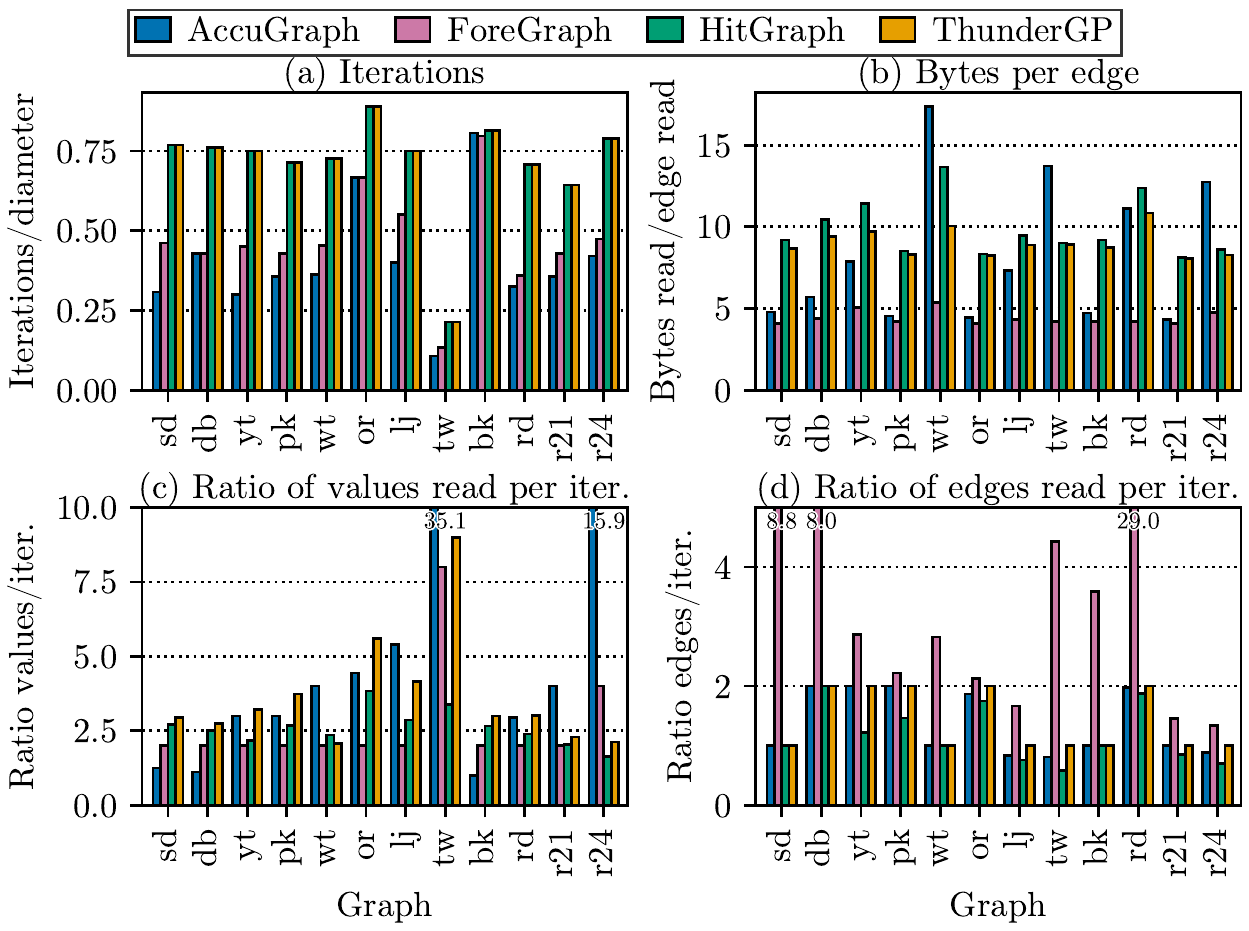}
	\vspace{-0.7cm}
	\caption{Critical performance metrics for BFS}
	\label{fig:aux}
\end{figure}
\cref{fig:comparison} shows a comparison of the four graph processing accelerators on all graphs from \cref{tab:graphs} for BFS, PR (one iteration), and WCC.
While we start with a general performance analysis with respect to accelerator design decisions and graph problems (performance dimensions (i) and (ii)) in this subsection, we refer back to \cref{fig:comparison} throughout the whole of \cref{sec:comparison} for more particular performance effects.
We also introduce four critical performance metrics in \cref{fig:aux} to explain our observations throughout \cref{sec:comparison}.
\emph{Number of iterations} has the biggest impact on performance since each iteration entails prefetching, edge, and value reading and writing (\cref{fig:aux}(a)).
Edge reading is the dominating factor of each iteration.
Thus, \emph{bytes per edge} (\cref{fig:aux}(b)) and \emph{edges read per iteration} (\cref{fig:aux}(d)) are the two second most important performance metrics.
Finally, \emph{values read per iteration} (\cref{fig:aux}(c)) can especially play a role for sparse and large graphs.
We only show BFS plots in \cref{fig:aux} for brevity because the numbers are very similar for PR and WCC.

Overall, performance of PR is the highest because only one iteration is performed (cf. \cref{fig:comparison}).
BFS and WCC performance is overall similar.
For BFS and WCC, performance on the bk and rd graph --- which we break out in separate plots with a difference y-axis scale --- is significantly lower.
This, however, is expected as bk and rd are graphs with a large diameter and thus require many more iterations to complete.
We notice that AccuGraph and ForeGraph on average perform better than HitGraph and ThunderGP.
To explain this, we additionally notice that AccuGraph and ForeGraph finish in significantly less iterations over the graph for BFS and WCC (not shown) than HitGraph and ThunderGP relative to the graph's diameter (cf. \cref{fig:aux}(a)).
This is possible due to the immediate update propagation scheme of AccuGraph and ForeGraph leading to convergence to the result in less iterations (\emph{insight 1}).
The iteration reduction of immediate update propagation is even more pronounced for WCC leading to a more pronounced performance advantage of AccuGraph and ForeGraph over HitGraph and ThunderGP for WCC when compared with BFS (cf. \cref{fig:comparison}).
Additionally, AccuGraph and ForeGraph read significantly less bytes per edge on average (cf. \cref{fig:aux}(b)).
This is due to the CSR data structure of AccuGraph and the compressed edges for ForeGraph (\emph{insight 2}).
For AccuGraph the exact number of bytes depends on the density of the graph which we discuss in \cref{sec:data-sets}.
ForeGraph always needs 4 bytes per edge, 2 bytes per vertex identifier, and some additional bytes to prefetch the value intervals.
HitGraph and ThunderGP need 8 bytes for each edge plus prefetching values and reading updates.

We also measured performance on weighted graphs for HitGraph and ThunderGP on SSSP and SpMV.
However, there were no significant differences in performance besides overall longer runtimes due to bigger edge size (because of edge weights) compared to BFS and PR respectively.
Thus, we do exclude SSSP and SpMV plots for brevity (but show runtime measurements in the appendix \cref{tab:weighted}).

\vspace{-0.1cm}
\subsection{Data Set Characteristics: Graph Properties}
\label{sec:data-sets}
\vspace{-0.1cm}
In this subsection, we discuss performance effects observable in \cref{fig:comparison} due to the graph properties (performance dimension (iii)) size ($|E|$), density ($D_{avg}$), and skewness of the degree distribution (as Pearson's moment coefficient of skewness $\mathbb{E}[(\frac{D - \mu}{\sigma})^3]$ with $D$ the degrees of the graph).
The first trend we notice is AccuGraph and ForeGraph performance decreasing relative to HitGraph and ThunderGP for large graphs like r24 and tw.
For the immediate update propagation scheme, destination vertex values need to be present when processing an edge which leads to loading these values many times instead of just once for update application (cf. \cref{fig:aux}(c)).
Thus, immediate update propagation leads to more value reads for large graphs (\emph{insight 3}).
Particular to AccuGraph, we see that performance is especially good for small graphs with only one partition ($|V| < 1,024,000$ for our configuration) such as sd, db, and bk.
AccuGraph saves vertex value reads for these graphs with skipping the prefetch step because the values are already in on-chip memory (cf. \cref{fig:aux}(c)).
However, for large graphs, AccuGraph still needs $n + 1$ CSR pointers for each partition leading to less savings in bytes per edge with horizontal partitioning (\emph{insight 4}).
HitGraph and ThunderGP performance is very similar in general.
We only see a significant difference in performance for the tw graph due to ThunderGP reading many more values because of vertical partitioning scheme.
HitGraph counteracts excessive value reads with an optimization described in \cref{sec:optimizations}.

\footnotetext{Raw performance numbers can be found in \cref{tab:raw} in \cref{sec:app1}}

All accelerator approaches benefit from dense graphs, with the effect being more pronounced for AccuGraph and ForeGraph (only working a full potential when $D_{avg} > 16$) due to a significant amount of pipeline stalls for sparse graphs.\footnote{Detailed performance by average degree can be found in \cref{fig:sparsity} in \cref{sec:app2}}
For accelerators with immediate update propagation and sparse graphs like db, yt, and rd, vertex value reads make up significantly more of the runtime (addition to \emph{insight 3}). 
Additionally, AccuGraph performance suffers for sparse graphs because the ratio between pointers and neighbors in the CSR data structure is higher.
\begin{figure}[bt]
	\centering
	\includegraphics[width=\linewidth]{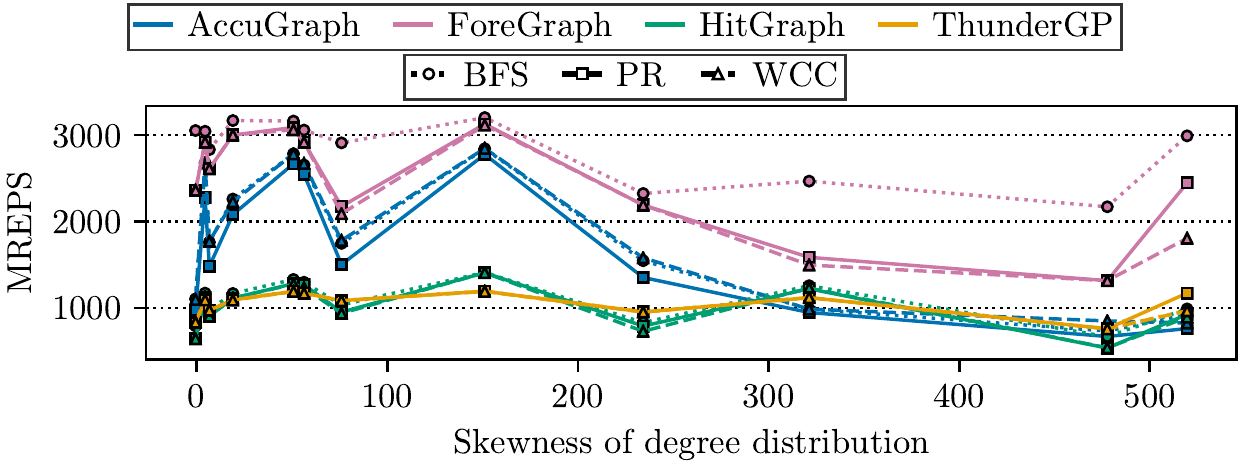}
	\vspace{-0.7cm}
	\caption{Performance by skewness of degree distribution}
	\label{fig:distribution}
\end{figure}
\cref{fig:distribution} shows raw edge processing performance by skewness of degree distribution.
Performance for AccuGraph and ForeGraph drops for graphs with high skewness (\eg wt and tw).
The accelerators are only working at their full potential at low to moderate skewness.
We also identify pipeline stalls caused by edge materialization on the CSR data structure as a problem for AccuGraph for high degree distribution skewness, similar to sparse graphs (\emph{insight 5}).
For ForeGraph we identify partition skew as the main cause of reduced performance which we discuss in more detail in \cref{sec:optimizations} but can be observed in \cref{fig:aux}(d).

\vspace{-0.1cm}
\subsection{Memory Technology: DRAM Types}
\label{sec:dram}
\vspace{-0.1cm}

\begin{figure}[bt]
	\centering
	\includegraphics[width=\linewidth]{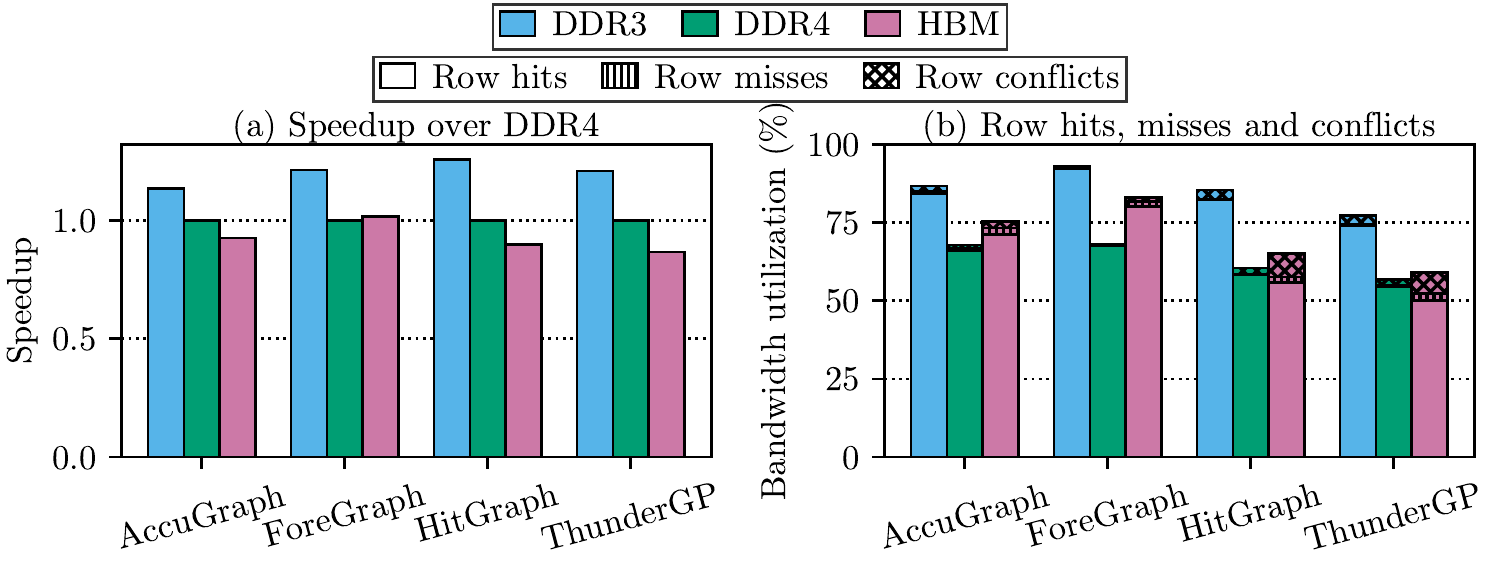}
	\vspace{-0.7cm}
	\caption{DDR3 and HBM speedup over DDR4 and bandwidth utilization with row hits, misses, and conflicts for BFS (single-channel)\protect\footnotemark}
	\label{fig:rows}
\end{figure}
\footnotetext{Raw performance numbers can be found in \cref{tab:ddr3} in \cref{sec:app1}}
In \cref{fig:rows}(a), we show average speedup of DRAM types (DDR3 and HBM) over DDR4 for all four accelerators (performance dimension (iv)).
We observe that modern memory (\eg DDR4 or HBM) does not necessarily perform better than the older DDR3, despite higher theoretical throughput.
This stems from lower bandwidth utilization and higher latency of requests (\emph{insight 6})
which is explained by extremely low utilization of parallelism in the memory due to mostly sequential reading of very few data structures at once.
AccuGraph and ForeGraph show more row hits due to write requests reusing rows in the row buffer of read requests (cf. \cref{fig:rows}(b)).
To achieve very good bandwidth utilization, approaches have to utilize either even more locality (ForeGraph and AccuGraph) or more memory parallelism (AccuGraph with its CSR data structure).
Additionally, there is no inherent benefit in using HBM for graph processing accelerators when the accelerator does not scale to multi-channel setups or scales poorly.
The bandwidth utilization goes up slightly for HBM (cf. \cref{fig:rows}(b)) but at a cost of significantly more latency inducing row misses and conflicts due to HBM's smaller row buffers.
Thus, we conclude that HitGraph and ThunderGP have higher potential adapting to HBM.

\begin{figure}[bt]
	\centering
	\includegraphics[width=\linewidth]{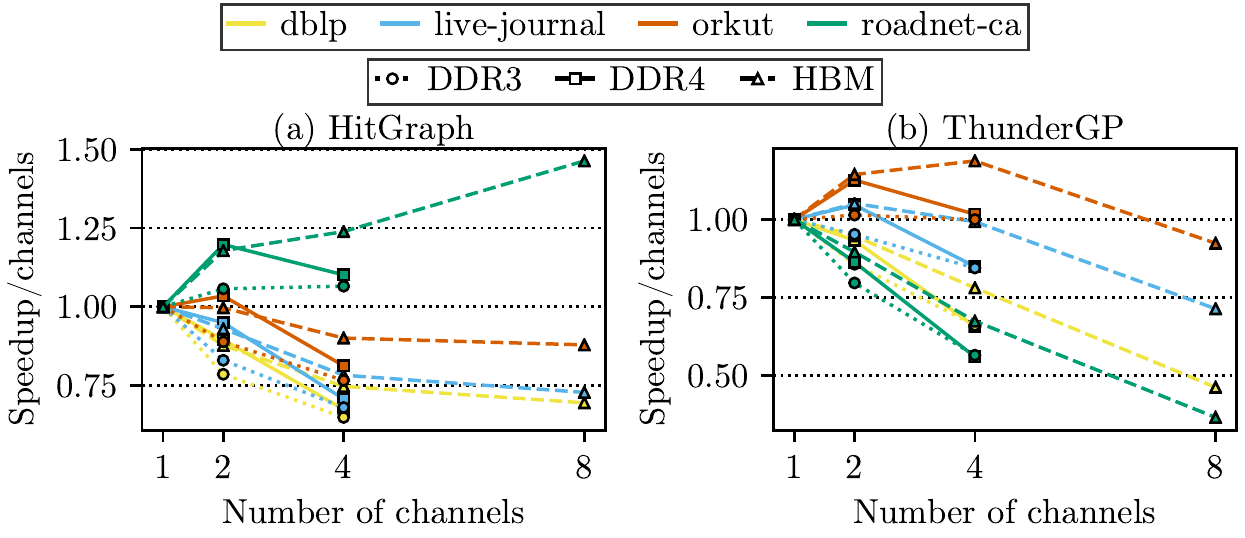}
	\vspace{-0.7cm}
	\caption{Scalability over number of channels for BFS\protect\footnotemark}
	\label{fig:scalability}
\end{figure}
\footnotetext{Raw performance numbers can be found in \cref{tab:scalability} in \cref{sec:app1}}
\Cref{fig:scalability} shows multi-channel scalability of HitGraph and ThunderGP.
AccuGraph and ForeGraph are not enabled for multi-channel operation and thus excluded.
For measurements on more than one channel we assume the clock frequency (reported by the respective paper) achieved with the 4-channel designs for both HitGraph and ThunderGP for the simulation environment.
However, as a limitation of these measurements this may not be exactly representative of the performance on real hardware because the clock speed could be slightly higher for two channels and lower for eight channels.
For HitGraph, we see almost linear performance improvements when increasing the number of channels and see super-linear improvements for the roadnet-ca graph.
This is due to improved effect of partition skipping resulting in significantly less requests to memory (\emph{insight 7}).
For ThunderGP, we see mostly sub-linear improvement in performance.
We explain the ThunderGP performance with its vertical partitioning scheme which leads to every PE working on values from all vertices such that all updates have to be applied to all channels, limiting performance (\emph{insight 8}).
However, the scaling seems to benefit from dense input graphs like the orkut graph.
An effect that we also observe is that DDR4 performance scales very well for two channels due to better bank parallelism utilization compared to DDR3 and HBM leading to better latency.

Another point we want to highlight is that HitGraph scales linearly in memory footprint with the number of memory channels and ThunderGP scales sub-linearly. 
HitGraph needs $n + m + n$ while ThunderGP requires $n*c + m + n*c$ space in memory where $n$ is the size of the vertex value array, $m$ is the size of the edge array and $c$ is the number of channels.
This not only means higher memory usage but also number of reads and writes not scaling linearly with number of channels for vertical partitioning (\emph{insight 9}).

\vspace{-0.1cm}
\subsection{Memory Access Optimizations}
\label{sec:optimizations}
\vspace{-0.1cm}
\begin{figure}[bt]
	\centering
	\includegraphics[width=\linewidth]{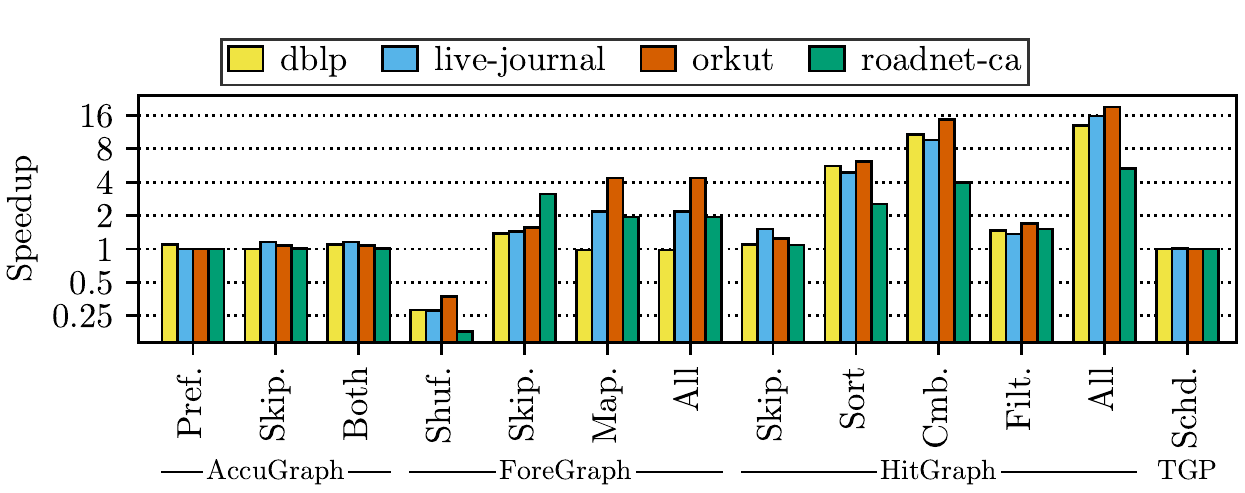}
	\vspace{-0.7cm}
	\caption{Optimization speedup for BFS (TGP: ThunderGP)\protect\footnotemark}
	\label{fig:optimizations}
\end{figure}
\footnotetext{Raw performance numbers can be found in \cref{tab:optimizations} in \cref{sec:app1}}
Each accelerator proposes a set of optimizations to reduce load on the memory or partition skew (performance dimension (v)).
In the following, we describe the different optimization approaches.
Previous \mbox{work \cite{DBLP:journals/corr/abs-2010-13619}} introduced the two optimizations prefetch (Pref.) and partition skipping (Skip.) for AccuGraph.
Prefetching is skipped when the on-chip partition is equal to the to-be-prefetched partition.
Partitions are skipped completely if none of their source values changed.
\Cref{fig:optimizations} shows a small improvement on the baseline at no significant added complexity.

ForeGraph shows bigger performance improvements for its three optimizations.
Edge shuffling (Shuf.) is a preprocessing step that repacks the edges such that the edge lists of $p$ shards are zipped into one (where $p$ is the number of PEs).
This alone leads to reduced performance due to aggravated load imbalance with partitions (due to padding in the form of null edges) but improves PE utilization when combined with stride mapping (Map.).
Shard skipping (Skip.) is employed for shards with unchanged source intervals compared to the previous iteration (equal to partition skipping for AccuGraph).
Finally, stride mapping renames vertices such that intervals are sets of vertices with a constant stride instead of consecutive vertices.
In total, the optimizations improve performance for all graphs we tested on (cf. \cref{fig:optimizations}).
However, we observe lower than average performance for sd, db, and rd.
For those graphs, among others, the interval-shard partitioning introduces a lot of partition skew (especially in combination with edge shuffling) leading to many more edges read than necessary to process the graph (cf. \cref{fig:aux}(d)).
Additionally, ForeGraph performance is higher for WCC on bk and rd when compared to BFS which is explained by less edges read due to more partition skipping (addition to \emph{insight 7}).

HitGraph, like AccuGraph and ForeGraph, employs partition skipping (Skip.) with similar effectiveness.
As a second optimization, HitGraph applies edge sorting by destination vertex (Sort), increasing locality to the gather phases value writing.
The edge sorting for HitGraph prepares the data structure for update combining (Cmb.).
Updates with the same destination vertex are combined into one in the shuffle phase which reduces the number of updates $u$ from $u = |E|$ to $u < |V| \times p$ with number of PEs $p$.
As a second optimization to update generation, a bitmap with cardinality $n$ in BRAM saves for each vertex if its value was changed in the last iteration.
This enables update filtering (Filt.) of updates from inactive vertices, saving a significant number of update writes.

ThunderGP proposes an offline scheduling of chunks to memory channels (Schd.) based on a heuristic predicting execution time.
Chunks are greedily scheduled such that the overall predicted execution time is as similar as possible.
Based on our measurements in \cref{fig:optimizations} this however does not make a big difference.
Additionally, ThunderGP does zero-degree vertex removal which we disabled for all runs because it is a pre-processing step applicable to all graph processing systems but hides performance ramifications of highly skewed degree distributions, \eg for wiki-talk or rmat-21-86.

\vspace{-0.1cm}
\subsection{Discussion}
\label{sec:discussion}
\vspace{-0.1cm}
In summary, from this comprehensive analysis of the four accelerators we gained nine insights that we categorize and group as trade-offs where possible in this section.
Our biggest finding is a trade-off between lower iteration count of immediate update propagation for graph problems like BFS and WCC when compared to 2-phase update propagation (\emph{insight 1}, similarly on CPU \cite{conf/europar/WhangLDP15}), and reading vertex values many more times for large graphs (\emph{insight 3}).
As an open challenge we propose finding an approach to reduce vertex value reads for immediate update propagation (\eg similar to update filtering for HitGraph) to lower the impact of graph size on the accelerator performance (\emph{open challenge (a)}).
As a second trade-off we found that CSR significantly reduces bytes per edge and values read for small and dense graphs (\emph{insight 2}) at a trade-off of reading more bytes per edge and values for large and sparse graphs when using horizontal partitioning (\emph{insight 4}).
As a last trade-off we found that large partitions reduce partition overhead while small partitions can significantly benefit partition skipping leading to super-linear performance increases (\emph{insight 7}).
We noticed that high skewness in degree distribution can lead to performance degradation, \eg for accelerators using CSR or interval-shard partitioning (\emph{insight 5}).
Additionally, vertical partitioning leads to poor channel scalability (\emph{insight 8}) and memory footprint for multi-channel setups (\emph{insight 9}).
Regarding modern memory (\eg HBM), we saw that trading of more latency with higher bank-level parallelism does not necessarily lead to better performance (\emph{insight 6}, generally for modern memory \cite{journals/corr/abs-1902-07609}).
Thus, we propose as an open challenge to investigate schemes to improve utilization of bank-level parallelism in modern memories (\emph{open challenge (b)}).
Lastly, we see and open challenge on enabling the immediate update propagation scheme for multi-channel (\emph{open challenge (c)}).

\vspace{-0.1cm}
\section{Related Work}
\label{sec:related_work}
\vspace{-0.1cm}
In \cite{journals/corr/abs-1902-07609}, the interactions of workloads and DRAM types is explored with Ramulator in great detail.
They study CPU-based workloads but do not cover FPGA accelerators.

\labeltitle{Accelerator simulation} \cite{conf/reconfig/ZhouCP15} introduces a simulation for HitGraph which generates the sequence of requests, but instead of simulating DRAM runtime, it assumes that every request results in a row buffer hit and models the performance along the cycles needed for processing the data and approximated pipelines stalls.
They, however, do not show performance numbers generated with this simulation.
\cite{conf/islped/YanHLALMDYZF019} uses Ramulator as the underlying DRAM simulator for a custom cycle-accurate simulation of Graphicionado \cite{conf/micro/HamWSSM16}.
However, this incurs very high implementation time.

\labeltitle{Modern memory technologies} In line with our findings on HBM, Schmidt et al. \cite{conf/memsys/SchmidtFB16} found that it is not trivial to attain good performance on another relatively new memory technology hybrid memory cube (HMC).
\cite{conf/fccm/WangHZA20} also confirms that it is crucial to consider how HBM should be used in FPGA-based accelerator designs.

\labeltitle{Trade-offs for cloud graph processing} Xu et al. benchmark different CPU-based graph processing systems in the cloud in \cite{conf/icws/XuZLSGWLZ17}.
For a limited set of graphs they study the performance of three systems.
They find that GridGraph \cite{conf/usenix/ZhuHC15}, which initially introduced the interval-shard partitioning, performs well, which we also observe for ForeGraph on single-channel systems.

\vspace{-0.1cm}
\section{Conclusion}
\label{sec:conclusion}
\vspace{-0.1cm}
This work addresses an important shortcoming of graph processing accelerators, namely \emph{comparability} of graph processing \emph{performance}.
We approach this matter by extending the DRAM-based simulation environment proposed in \cite{DBLP:journals/corr/abs-2010-13619} and compare the performance of four well-known graph accelerators (\ie AccuGraph, ForeGraph, HitGraph, and ThunderGP) along performance dimensions relevant for graph processing (cf. dimensions (i)--(v)).
We found performance effects based on accelerator design decisions (\emph{insights 5, 8, 9}), issues in utilization of modern memory technologies (\emph{insight 6}), and several interesting trade-offs (\emph{insights 1--4, 7}).

We propose to conduct future work on the identified \emph{open challenges (a)--(c)}, \ie further improving the immediate update propagation scheme for large graphs, leveraging the potential of HBM for graph processing, and multi-channel scalability of the immediate update propagation scheme.
Additionally, we see a need for standardization of benchmark techniques in the field of graph processing accelerators, as sketched in this work.

\vspace{-0.1cm}
\bibliographystyle{ACM-Reference-Format}
\bibliography{paper}

\newpage
\appendix

\vspace{-0.1cm}
\section{Raw data}
\label{sec:app1}
\vspace{-0.1cm}
{\setlength{\tabcolsep}{0.275em}
\begin{table}[H]
\tiny
\centering
\begin{tabular}{l r r r r r r r r r r r r}
    & \multicolumn{3}{c}{AccuGraph} & \multicolumn{3}{c}{ForeGraph} & \multicolumn{3}{c}{HitGraph} & \multicolumn{3}{c}{ThunderGP} \\
    G & BFS & PR & WCC & BFS & PR & WCC & BFS & PR & WCC & BFS & PR & WCC \\
    \hline
	sd & 0.0017 & 0.0005 & 0.0009 & 0.0159 & 0.0009 & 0.0046 & 0.0081 & 0.0009 & 0.0077 & 0.0087 & 0.0009 & 0.0078 \\
	db & 0.0107 & 0.0014 & 0.0083 & 0.0268 & 0.0019 & 0.0173 & 0.0344 & 0.0023 & 0.0348 & 0.0345 & 0.0022 & 0.0323 \\
	yt & 0.0232 & 0.0044 & 0.0189 & 0.0332 & 0.0032 & 0.0256 & 0.0659 & 0.0076 & 0.0706 & 0.0940 & 0.0063 & 0.0879 \\
	pk & 0.1154 & 0.0241 & 0.0688 & 0.1335 & 0.0225 & 0.1126 & 0.3465 & 0.0484 & 0.3310 & 0.5225 & 0.0523 & 0.5239 \\
	wt & 0.0274 & 0.0075 & 0.0236 & 0.0327 & 0.0061 & 0.0245 & 0.0601 & 0.0094 & 0.0653 & 0.0529 & 0.0066 & 0.0464 \\
	or & 0.4709 & 0.0879 & 0.1685 & 0.4736 & 0.0791 & 0.2791 & 1.2344 & 0.1831 & 1.2852 & 1.5718 & 0.1967 & 1.5754 \\
	lj & 0.2650 & 0.0459 & 0.2202 & 0.4347 & 0.0396 & 0.2577 & 0.7591 & 0.0725 & 0.9049 & 0.9538 & 0.0637 & 0.9555 \\
	tw & 10.3114 & 1.9304 & 10.4346 & 21.7350 & 2.7537 & 63.8956 & 13.8804 & 1.5886 & 20.0293 & 24.2738 & 1.2539 & 66.8212 \\
	bk & 1.6355 & 0.0033 & 1.6219 & 5.0959 & 0.0057 & 3.2011 & 3.7714 & 0.0068 & 4.7490 & 4.0371 & 0.0070 & 4.8985 \\
	rd & 1.3653 & 0.0057 & 0.9357 & 8.0324 & 0.0108 & 2.7803 & 3.9504 & 0.0086 & 4.6874 & 4.0059 & 0.0067 & 3.6763 \\
	r21 & 0.3174 & 0.0650 & 0.3466 & 0.4926 & 0.0681 & 0.3757 & 0.9812 & 0.1282 & 1.2820 & 1.3596 & 0.1512 & 1.5147 \\
	r24 & 1.9207 & 0.2835 & 1.8342 & 1.3074 & 0.2287 & 1.5206 & 2.2484 & 0.2198 & 2.7620 & 3.5936 & 0.2401 & 3.3590 \\
\end{tabular}

\medskip
G: Graph

\caption{DDR4 (single-channel) runtime measurements (in seconds) with all optimizations enabled}
\label{tab:raw}
\end{table}}
\vspace{-0.7cm}
\Cref{tab:raw} shows raw performance results for single-channel DDR4 simulations of AccuGraph, ForeGraph, HitGraph, and ThunderGP as runtime in seconds.
\Cref{fig:comparison} shows the derivative performance metric MTEPS based on the raw data in \cref{tab:raw} which is attained by dividing $|E|$ from \cref{tab:graphs} through the runtime.

{\setlength{\tabcolsep}{0.275em}
\begin{table}[H]
\footnotesize
\centering
\begin{tabular}{l r r r r}
    & \multicolumn{2}{c}{HitGraph} & \multicolumn{2}{c}{ThunderGP} \\
    Graph & SSSP & SpMV & SSSP & SpMV \\
    \hline
	sd & 0.0114 & 0.0012 & 0.0122 & 0.0012 \\
    db & 0.0459 & 0.0030 & 0.0469 & 0.0029 \\
    yt & 0.0848 & 0.0096 & 0.1271 & 0.0084 \\
    pk & 0.5014 & 0.0695 & 0.7501 & 0.0747 \\
    wt & 0.0740 & 0.0111 & 0.0680 & 0.0085 \\
    or & 1.8002 & 0.2639 & 2.2647 & 0.2821 \\
    lj & 1.0300 & 0.0964 & 1.3311 & 0.0884 \\
    tw & 18.6132 & 2.0955 & 32.4852 & 2.0255 \\
    bk & 5.2940 & 0.0094 & 5.6896 & 0.0098 \\
    rd & 5.0307 & 0.0105 & 5.1446 & 0.0085 \\
    r21 & 1.4582 & 0.1904 & 1.9629 & 0.2173 \\
    r24 & 3.2229 & 0.3124 & 5.0438 & 0.3355 \\
\end{tabular}

\caption{Weighted graph runtime measurements (in seconds) with all optimizations enabled on DDR4 (single-channel)}
\label{tab:weighted}
\end{table}}
\vspace{-0.7cm}
\cref{tab:weighted} shows raw performance results for single-channel DDR4 simulations of HitGraph and ThunderGP on weighted graphs.
AccuGraph and ForeGraph do not support weighted graphs and are thus excluded.

{\setlength{\tabcolsep}{0.275em}
\begin{table}[H]
\footnotesize
\centering
\begin{tabular}{l r r r r r r r r}
     & \multicolumn{2}{c}{AccuGraph} & \multicolumn{2}{c}{ForeGraph} & \multicolumn{2}{c}{HitGraph} & \multicolumn{2}{c}{ThunderGP} \\
    Graph & DDR3 & HBM & DDR3 & HBM & DDR3 & HBM & DDR3 & HBM \\
    \hline
	sd & 0.0014 & 0.0017 & 0.0131 & 0.0157 & 0.0064 & 0.0090 & 0.0070 & 0.0096 \\
    db & 0.0094 & 0.0114 & 0.0221 & 0.0264 & 0.0273 & 0.0382 & 0.0289 & 0.0401 \\
    yt & 0.0200 & 0.0244 & 0.0274 & 0.0327 & 0.0526 & 0.0736 & 0.0769 & 0.1060 \\
    pk & 0.0970 & 0.1157 & 0.1101 & 0.1316 & 0.0275 & 0.0389 & 0.4261 & 0.5833 \\
    wt & 0.0241 & 0.0303 & 0.0269 & 0.0321 & 0.0484 & 0.0671 & 0.0422 & 0.0576 \\
    or & 0.3935 & 0.4708 & 0.3905 & 0.4668 & 0.9660 & 1.3605 & 1.2889 & 1.7739 \\
    lj & 0.2335 & 0.2867 & 0.3584 & 0.4282 & 0.6045 & 0.8461 & 0.7893 & 1.1007 \\
    tw & 9.0370 & 11.2454 & 17.9232 & 21.4115 & 11.4310 & 16.3588 & 20.8722 & 30.920
1 \\
    bk & 1.3712 & 1.6510 & 4.2011 & 5.0245 & 2.9800 & 4.1829 & 3.3493 & 4.5960 \\
    rd & 1.1917 & 1.4289 & 6.6240 & 7.9176 & 3.1720 & 4.4374 & 3.3688 & 4.7319 \\
    r21 & 0.2651 & 0.3168 & 0.4062 & 0.4856 & 0.7626 & 1.0785 & 1.1087 & 1.5177 \\
    r24 & 1.6698 & 2.2024 & 1.0779 & 1.2862 & 1.7598 & 2.4812 & 3.0170 & 4.1784 \\
\end{tabular}

\caption{DDR3 and HBM (single-channel) runtime measurements (in seconds) with all optimizations enabled for BFS}
\label{tab:ddr3}
\end{table}}
\vspace{-0.7cm}
\cref{tab:ddr3} shows raw performance results for single-channel DDR3 and HBM simulations of AccuGraph, ForeGraph, HitGraph, and ThunderGP.
These numbers are presented as speedup over DDR4 in \cref{fig:rows}.
The respective DDR4 baseline is taken from \cref{tab:raw}.

{\setlength{\tabcolsep}{0.275em}
\begin{table}[H]
\footnotesize
\centering
\begin{tabular}{l r r r r r r r r r}
     &  & \multicolumn{4}{c}{HitGraph} & \multicolumn{4}{c}{ThunderGP} \\
    DRAM & \#Channels & db & lj & or & rd & db & lj & or & rd \\
    \hline
    \hline
    \multirow{2}{*}{DDR3} & 2 & 0.0174 & 0.3640 & 0.5433 & 1.5002 & 0.0169 & 0.4143 & 0.6355 & 2.1135 \\
     & 4 & 0.0105 & 0.2221 & 0.3151 & 0.7443 & 0.0109 & 0.2336 & 0.3222 & 1.4887 \\
    \hline
	\multirow{2}{*}{DDR4} & 2 & 0.0192 & 0.3998 & 0.5966 & 1.6494 & 0.0185 & 0.4557 & 0.6978 & 2.3198 \\
     & 4 & 0.0127 & 0.2682 & 0.3798 & 0.8968 & 0.0131 & 0.2807 & 0.3865 & 1.7867 \\
    \hline
    \multirow{3}{*}{HBM} & 2 & 0.0218 & 0.4549 & 0.6824 & 1.8830 & 0.0211 & 0.5236 & 0.7753 & 2.6404 \\
     & 4 & 0.0128 & 0.2702 & 0.3776 & 0.8957 & 0.0128 & 0.2772 & 0.3735 & 1.7533 \\
     & 8 & 0.0069 & 0.1452 & 0.1934 & 0.3792 & 0.0108 & 0.1926 & 0.2400 & 1.6126 \\
\end{tabular}

\caption{Multi-channel scalability runtime measurements (in seconds) with all optimizations enabled for BFS}
\label{tab:scalability}
\end{table}}
\vspace{-0.7cm}
\cref{tab:scalability} shows raw performance results for multi-channel simulations of HitGraph and ThunderGP with BFS on DDR3, DDR4, and HBM.
AccuGraph and ForeGraph do not support multi-channel memory and are thus excluded.
These numbers are presented as speedup over the respective single-channel simulation in \cref{fig:scalability}.
The single-channel baselines are taken from \cref{tab:raw} and \cref{tab:ddr3}.

{\setlength{\tabcolsep}{0.275em}
\begin{table}[H]
\footnotesize
\centering
\begin{tabular}{l l r r r r}
     &  & \multicolumn{4}{c}{Graph} \\
    Accelerator & Optimization & db & lj & or & rd \\
    \hline
    \hline
    \multirow{3}{*}{AccuGraph} & None & 0.0118 & 0.3062 & 0.5071 & 1.3834 \\
     & Prefetch skipping & 0.0107 & 0.3062 & 0.5071 & 1.3834 \\
     & Partition skipping & 0.0118 & 0.2650 & 0.4709 & 1.3670 \\
    \hline
    \multirow{4}{*}{ForeGraph} & None & 0.0263 & 0.9428 & 2.0590 & 15.6424 \\
     & Edge shuffling & 0.0936 & 3.3837 & 5.5188 & 86.4302 \\
     & Shard skipping & 0.0191 & 0.6594 & 1.3149 & 4.9896 \\
     & Stride mapping & 0.0268 & 0.4347 & 0.4736 & 8.0324 \\
    \hline
    \multirow{5}{*}{HitGraph} & None & 0.1594 & 4.1306 & 7.1937 & 4.7238 \\
     & Partition skipping & 0.1455 & 2.7382 & 5.8026 & 4.3559 \\
     & Edge sorting & 0.0284 & 0.8422 & 1.1732 & 1.8639 \\
     & Update combining & 0.0149 & 0.4318 & 0.4883 & 1.1849 \\
     & Update filtering & 0.1081 & 3.0243 & 4.2361 & 3.1239 \\
    \hline
    ThunderGP & None & 0.0125 & 0.2702 & 0.3701 & 1.7121 \\
\end{tabular}

\caption{Runtime measurements (in seconds) with different optimizations enabled for BFS}
\label{tab:optimizations}
\end{table}}
\vspace{-0.7cm}
\cref{tab:optimizations} shows raw performance results for AccuGraph, ForeGraph, HitGraph, and ThunderGP with all optimizations disabled and partially enabled optimizations for BFS on single-channel DDR4.
These numbers in addition with the corresponding numbers from \cref{tab:raw} (all optimizations enabled) are presented as speedup over the unoptimized runtimes in \cref{fig:optimizations}.

\vspace{-0.1cm}
\section{Edge Processing Performance}
\label{sec:app2}
\vspace{-0.1cm}
\begin{figure}[H]
	\centering
	\includegraphics[width=\linewidth]{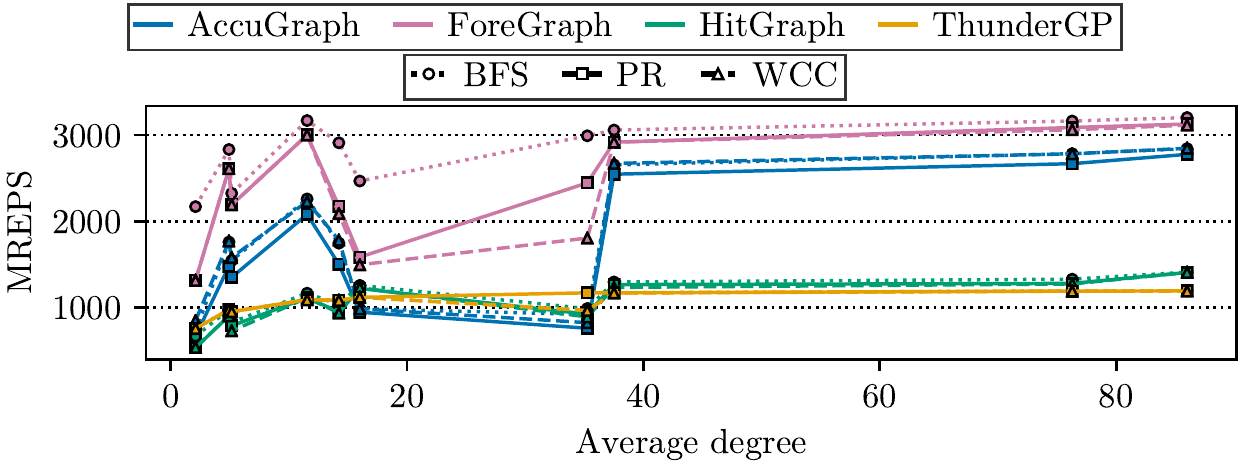}
	\vspace{-0.7cm}
	\caption{Performance by average degree of the graph}
	\label{fig:sparsity}
\end{figure}
\Cref{fig:sparsity} shows raw edge processing performance in MREPS by average degree of the graph ($D_{avg}$ in \cref{tab:graphs}).

\end{document}